\documentclass[]{rmaa}

\usepackage[english]{babel}
\usepackage{amsmath}
\usepackage{url}

\newcommand{\VSp}[1]{VVV-SkZ\_pi\-pe\-li\-ne}
\newcommand{\DoP}[1]{Do\-Phot}
\newcommand{\DAOP}[1]{DAO\-PHOT}
\newcommand{\ALS}[1]{ALL\-STAR}
\newcommand{\ALF}[1]{ALL\-FRAME}
\newcommand{\DAOM}[1]{DAO\-MA\-STER}

\newcommand{\Ks}[1]{K_{s #1}}

\newcommand{\degr}{^\circ}
\newcommand{\hgd}{\textsf{high good datum}}

\title{The \VSp{}: an automatic PSF-fitting photometric pipeline\\ for the VVV survey}

\author{Francesco Mauro\altaffilmark{1}, Christian Moni Bidin\altaffilmark{1,2,5}, Andr\'e-Nicolas Chen\'e\altaffilmark{1,3}, Doug Geisler\altaffilmark{1}, Javier Alonso-Garc\'{\i}a\altaffilmark{4}, Jura Borissova\altaffilmark{3,5}, Giovanni Carraro\altaffilmark{6,7}}
\altaffiltext{1}{Departamento de Astronom\'ia, Universidad de Concepci\'on, Casilla 160-C, Concepci\'on, Chile}
\altaffiltext{2}{Instituto de Astronom\'ia, Universidad Cat\'olica del Norte, Av. Angamos 0610, Antofagasta, Chile}
\altaffiltext{3}{Departamento de de Fis\'ica y Astronom\'ia, Universidad de Valpara\'iso, Av. Gran Breta\~na 1111, Playa Ancha, Casilla 5030, Chile}
\altaffiltext{4}{Departamento de Astronom\'ia y Astrof\'isica, Pontificia Universidad Cat\'olica de Chile, Casilla 306, Santiago, Chile}
\altaffiltext{5}{The Milky Way Millennium Nucleus, Av. Vicu\~{n}a Mackenna 4860, 782-0436 Macul, Santiago, Chile}
\altaffiltext{6}{European Southern Observatory, Ave. Alonso de Cordova 3107, Casilla 19, 19001 Santiago, Chile}
\altaffiltext{7}{Dipartimento di Fisica e Astronomia, Universit\`a di Padova, Via Marzolo 8, I-35131 Padua, Italy}

\shorttitle{\VSp{}}
\shortauthor{Mauro et~al.}

\abstract{
We present the \VSp{}, a \DAOP{}-based photometric pipeline, created to perform PSF-fitting photometry of  ``VISTA Variables in the V\'ia L\'actea'' (VVV) ESO Public Survey data.
The pipeline replaces the user avoiding  repetitive interaction in all the operations, retaining all of the benefits of the power and accuracy of the \DAOP{} suite.
The pipeline provides an astrometrized photometric catalog reliable up to more than 2 magnitudes brighter than the saturation limit, where other techniques fail.
It also produces deeper and more accurate photometry.
These achievements allow the \VSp{} to produce data well anchored to the selected standard photometric system and analyze important phenomena (i.e. TRGB, RGB slope, HB morphology, RR Lyrae), that other methods are not able to manage.
}
\resumen{
Presentamos la \VSp{}, una pipeline fotom\'etrica basada en la suite \DAOP{}, creada para realizar fotometr\'{\i}a PSF de los datos de la sondeo p\'ublico de la ESO ``VISTA Variables in the V\'ia L\'actea'' (VVV).
La pipeline reemplaza el usuario en el uso iterativo y repetitivo en todas las operaciones, manteniendo todos los beneficios de la precisi\'on de la suite \DAOP{}.
La pipeline suministra un catalogo fotom\'etrico astrometrizado confiable hasta regiones m\'as de 2 magnitudes m\'as brillante que el limite de saturaci\'on.
La pipeline tambi\'en produce fotometr\'{\i}a m\'as profunda y m\'as precisa.
Estos logros permiten a la pipeline \VSp{} producir resultados bien anclados al sistema fotom\'etrico estandard selecionado, y analizar regiones muy importante (como TRGB, la pendiente de la RGB, morfolog\'{\i}a de la HB, las estrellas RR Lyrae) que otros m\'etodos no alcanzan.
}

\addkeyword{techniques: photometric}

\begin{document}
\maketitle

\section{Introduction}
\label{s_intro}

In the last decade, large-area surveys have revolutionized our understanding of the Galaxy and the Universe.
Surveys like the Sloan Digital Sky Survey \citep{SDSS} and the Two Micron All Sky Survey (2MASS) \citep{2MASS} can be defined as two of ``the most ambitious and influential surveys in the history of astronomy''\footnote{\url{http://www.sdss.org/}}, for their contribution to astronomical knowledge.
The importance of surveys in astronomical studies is underlined by the recent development of several survey-dedicated telescopes, like the 4-meter Visible and Infrared Survey Telescope for Astronomy (VISTA), the 2.5m VLT Survey Telescope, and the future 8.4m Large Synoptic Survey Telescope.

The ``VISTA Variables in the V\'ia L\'actea'' (VVV) Survey  \citep{Minniti10,Saito10} is one of the six ESO Public Surveys operating on VISTA.
VVV scans the Galactic bulge ($-10\leq l\leq+10$, $-10\leq b\leq +5$) and an adjacent part of the southern disk ($-65\leq l\leq-10$, $-2\leq b\leq +2$) in five near-IR bands ($ZYJHKs{}$) with the VIRCAM camera \citep{Emerson10}, an array of sixteen non-buttable 2048$\times$2048~pixel detectors, with a pixel scale of $\sim 0\farcs 34/pix$.
The VVV data are organized in contiguous $1\degr\times1.5\degr$ areas called ``tiles''.
Because of the large gaps among the CCDs, six offset exposures, called ``pawprints''\footnote{\url{http://www.vista.ac.uk/}}, are required to survey a tile area covering each point at least twice.
Each pawprint, on the other hand, is the combination of two exposures jittered by about 60 pixels in each axis.
Multiple additional $2\times 4$s $K_s$-band images are obtained each year to find and monitor the variable stars in the VVV area.
Observations are separated in observation blocks (OBs), that consist of exposures in $JHKs{}$, or $ZY$, or $2\times 4$s $Ks{}$ exposures of a single tile.
VVV data are automatically processed at the Cambridge Astronomical Survey Unit (CASU)\footnote{http://casu.ast.cam.ac.uk/} with the VIRCAM pipeline \citep{Irwin04}, and the catalogs of aperture photometry (hereafter CASU catalogs) are produced and made publicly available through the ESO archive\footnote{\url{http://www.eso.org/sci/archive.html} ~ \url{http://archive.eso.org/wdb/wdb/adp/phase3_main/form}} \citep{Saito12}.

The VVV survey presents many situations in which the results of aperture photometry are clearly unsatisfactory, such as the very crowded regions of the Galactic center or stellar clusters, and the small-scale spatial variations of the background in the direction of star forming regions in the Galactic disk.
In these cases, PSF-fitting photometry is mandatory to obtain optimal results.
Moreover,  aperture photometry does not return reliable results for saturated stars ($\Ks{}< 12$ for $2\times 4$s exposures).
This prohibits the study of a variety of important objects (e.g. most of the red giant branch and RR Lyrae stars of bright globular clusters, or OB variables in open clusters).
Fortunately, the wings of the PSF of these stars, unaffected by saturation, can be used to derive reliable PSF-fitting photometry.

Performing PSF-fitting photometry on multiple offsets in five bands requires a large amount of CPU time and also valuable interaction time to select optimal parameters.
In addition, the image distortions \citep{Saito12} complicate the generation of stacked images to obtain a master list of stars when different chips or more than two offsets of the same chip are involved.
The image distortions are up to about 10--15\% across the wide field of view, and require a native WCS distortion model for the pawprints that varies radially as a fifth order polynomial.

In this paper, we present the \VSp{} (VSp), an automated pipeline designed to produce optimal PSF-fitting photometry with VVV data using the \DAOP{} suite \citep{DAOPHOT}.
VSp is an adaptation of a previous procedure, developed by one of us (FM) to deal with data from a generic telescope.
Its aims were to minimize the problem of the high degree of  repetitive interaction required when \DAOP{} is used on a large amount of data, to produce accurate, precise and uniform photometry, and to leave to the user only the selection of the parameters.
To achieve this result, algorithms were developed to manage a number of  important tasks, like the rejection of the worst PSF stars.
The software described here has already been applied in several papers with excellent results \citep[e.g.,][]{MoniBidin2011,Majaess11,Majaess12,MauroAS,Mauro2012,MauroPhD,BaumeAS,BorissovaAS,CheneAS,Chene2012,Chene2012b}, where the reader can check the quality of the software products in different astrophysical contexts.
The original pipeline has been applied to the ``Southern Open Cluster Survey'' (SOCS) project \citep{Kinemuchi2010,MauroMSc}.

In Section \ref{s_general} we give a general description of the \VSp{} and its main characteristics.
In Section \ref{s_det} we discuss the pipeline in more detail.
In Section \ref{s_comp} we show the advantages of VSp as a source of VVV photometry, discussing the comparison between the catalog produced by VSp, the 2MASS catalog (that was used for calibration) and the catalogs of other procedures available to obtain photometry from the VVV survey data.

\section{General description}
\label{s_general}

The VSp performs PSF-fitting photometry through standard \DAOP{}\textsc{iv} and \ALF{} routines \citep[][manuals and private communication]{ALLFRAME}, called by a series of Perl\footnote{www.perl.org} \citep{perl} scripts, that use programs written in C code with double precision  when mathematical accuracy is required.
The pipeline is composed of six parts, each one operated by a different script, whose input depends only on the output of the previous script.
This design permits one to run the parts of the pipeline in series or separately, or to re-run the procedure from any point.
The software is designed to perform photometric calibration using a generic catalog with standard $JH\Ks{}$ magnitudes, but also with $Z$ and $Y$, if available.
The astrometry is based on the World Coordinate System (WCS) present in the pawprints, determined by CASU and based on 2MASS astrometry.
The pipeline depends on some external programs for specialized tasks (i.e., gnuplot for the plots, IRAF to extract images and header information; see Appendix \ref{ap_progr}).
The pipeline is designed to work on the uncompressed pawprints, pre-reduced by CASU.
In addition to the images, the required input consists of a list of frames, the configuration file of the pipeline, and a catalog of calibration stars with their equatorial coordinates and at least their standard $JH\Ks{}$ magnitudes.
The list of frames must contain the input file name, the number of the chip whose image has to be extracted, and the output file name (see details in Section \ref{ss_prep}).
The default values in the option file, which governs the main parameters of the pipeline, have been extensively tested and should work well in most cases.
However, the values of the parameters are tunable by the user.
The only mandatory input information is related to the standard star catalog, since all the other information (such as seeing, filter name, etc) is obtained automatically from the image headers.
The pipeline is characterized by a comprehensive procedure for the PSF calculation, including the possibility to give an optional list of integer coordinates of rectangular regions, whose stars are to be excluded from the PSF calculation.
The user could desire to avoid stars in areas not suitable for the PSF calculation (i.e., variable background, highly saturated stars,  highly crowded regions, defects of the image).
This list must be given for each image that needs it.
Additionally, any PSF star lying in the 60 pixel border area, covered by only one of the two jitters, is automatically rejected.

An important added value of the pipeline is the determination of the \DAOP{} parameter \hgd{}.
Fixing it requires particular care, because its accurate determination permits  good photometry even for partially saturated stars.
There is no unique value suitable for all the VVV images, because the saturation level varies with chip, filter, and exposure time.
Therefore, the pipeline reads the value to be used from an internal table, that provides the \hgd{} as a function of chip, filter, and OB type\footnote{The OBs are divided depending on the filters used (suffix j for the observations in $JHKs{}$, z for the observations in $ZY$ and v-$n$ for the additional epochs in $K_\mathrm{s}$), and whether they are bulge or disk tiles (prefix b and d).
The $JHKs{}$ bulge OBs have a different exposure time than disk OBs and, even when bulge and disk OBs have the same exposure time, we find that it is better to consider them separately.}.
The values were determined from an accurate statistical analysis of 48 pawprints.
We found a strong correlation  ($r_{xy}=0.97-0.99$) between the saturation level of different filter-OB pairs of the same chip, which helps simplify the internal table.
Taking into account their variations (across the chip or between different exposures) we fixed the \hgd{} at 80\% of the saturation level.
To avoid problems with exposures whose sky value is particularly high (like $H$ exposures in the disk), we decided to use the following equation 
\begin{equation}\label{eq:hgd}
 \mathsf{high\, good\, datum}=sky+0.80(SatLvl-sky) 
\end{equation}
where $sky$ is the sky value, and $SatLvl$ the saturation level.
The success of this determination is demonstrated by the comparison with external catalogs, like 2MASS (see Section \ref{ss_comp2MASS}).
Of course, the user can still provide the pipeline with their own values of \textsf{gain}, \textsf{read-out noise} and \hgd{}.

Once the pipeline has achieved a first measurement of the photometry, it stacks all the frames to create the master list of sources.
This is a huge advantage over getting a list for every pawprint, since it provides much deeper photometry before source detection.
The benefits of stacking different ``stripes''\footnote{A tile can be seen as the union of eight horizontal  $\sim 2105$~pixel-wide ``stripes'' (combination of twelve frames), with only $\sim 150$ pixels of overlap, four fifths of which are covered by only one jitter for each stripe.} are negligible, because of the very limited overlap.
Hence, the pipeline by default combines only the images of the same ``stripe'' collected with the same chip (note that other choices can be selected in the option file, e.g. the stacking of all the frames of the same ``stripe'', or separating them by offset).
The combination of frames collected with the same chip, or with the central chips of the camera array, is generally not a problem.
However, distortion across the wide field of the VIRCAM camera completely spoils the results when using widely-separated chips, even when \DAOM{} is used with cubic transformations, and combining frames from different chips is often problematic.

The final \ALF{} output is astrometrized before matching the single output catalogs, to avoid the problems introduced by the field distortion.
Consequently, their matching is simplified.
The pipeline transforms the frame position in pixels into relative positions ($\cos{\delta}\Delta \alpha;\Delta \delta$) in arcseconds.
The origin is centered on the point with the furthest south and west values, respectively in right ascension and declination, in order to have positive coordinates for all the stars.
When the photometry in all the passbands are merged in the final catalog, the position in the equatorial system is calculated for each source.
The relative coordinate system is also included in the final catalog (see Appendix \ref{ap_Ctlg}), in addition to the positions in equatorial coordinates.

Another important added value of the pipeline is the spurious-detection cleaning procedure.
It is commonly known that saturated stars cause a non-negligible quantity of false star detections, that can adversely affect the results.
A cleaning process, composed of two independent procedures, is included in the code for this reason.
The first procedure is based on the trend with magnitude of the maxima of the distribution of photometric error $\sigma_m$, while the second procedure is based on both the typical loci of the spurious detections in the magnitude-error diagram, and their characteristic clustering around the saturated stars (see Section \ref{ss_metrcalib} for more details).

The source of the standard stars data is not constrained, but the procedure was tuned on 2MASS data and tested only with them.
In particular, the coordinates of the observed stars are derived in the 2MASS astrometric system, hence a successful cross-identification with another input catalog is not guaranteed in the presence of an astrometric offset with respect to 2MASS.
Feeding the pipeline with the complete list of standard stars in the field is recommended, as the code automatically selects the best stars, based on the magnitude, the quality of the catalog photometry, and the presence of contaminating nearby stars.
When using 2MASS PSC as the reference catalog, the optimal range lies between the highly-saturated star regime ($\Ks{}=8-10$) and the 2MASS deviations at the fainter end \citep[$\Ks{}=12-14$, see][]{MoniBidin2011}, both varying between fields.
It is therefore highly recommended to initially adopt the complete catalog with no pre-selection in magnitudes, then check the results and adjust the magnitude intervals, and  rerun the final script.
To facilitate this process, an astro-photometric comparison with the input standard catalog is executed at the end, the results are stored in a file, and a comprehensive set of graphs is generated to check the differences in both colors and magnitudes.
The difference in position with the 2MASS catalog generally have a sample standard deviation less than 0.2 arcsec.

The final output catalog contains the calibrated magnitudes only in the $JH\Ks{}$ bands.
The 2MASS catalog does not provide $Y$ and $Z$  magnitudes, and there is a general lack of standard stars in these bands suitable for calibration in the survey area.
We chose to leave these magnitudes in the instrumental system, since good calibration equations are not publicly available, or they depend on the spectral type.
The equatorial position resolution in the final catalog is  $10^{-6}$ degrees, which is the same resolution provided by the VizieR system\footnote{\url{http://vizier.u-strasbg.fr/}} and in other catalogs.

While the calibrated catalog is the final goal of the photometric work, the pipeline output also includes other auxiliary files, primarily aimed to check the quality of the results.
In addition to the aforementioned astro-photometric comparison with the standard catalog, the plots of the calibration equations, and the plots related to the cleaning procedure, the code also produces a stellar density map in the format of a fits image with included WCS, and some additional useful graphics, e.g. a color-magnitude diagram $(J-\Ks{}; \Ks{})$, a reddening-free color-magnitude diagram $(c_3;\Ks{})$ \citep[see ][]{Catelan11}, and a color-color diagram $(H-\Ks{}; J-H)$.

\section{Salient details}
\label{s_det}
In this Section we give more relevant details of the main parts of the \VSp{}, and briefly describe how they operate.
The fundamental steps of the reduction are: the creation of the input files, an accurate determination of the PSF, the creation of a complete master list of stars, the PSF-fitting photometry with \ALF{}, and, finally, the astrometrization, calibration, and matching of all the data.
We will use the convention that the script and program names will be in typewriter (e.g., \texttt{VVV-GetImg\-Info\-Hdr.pl}), the file names  in italics (e.g., \emph{\VSp{}.opt}), and the options in sans serif (e.g, \textsf{stdcat}).
Figure 1 shows a flow chart of how the program works.

\begin{figure}[!ht]
\centering
\includegraphics[scale=.5]{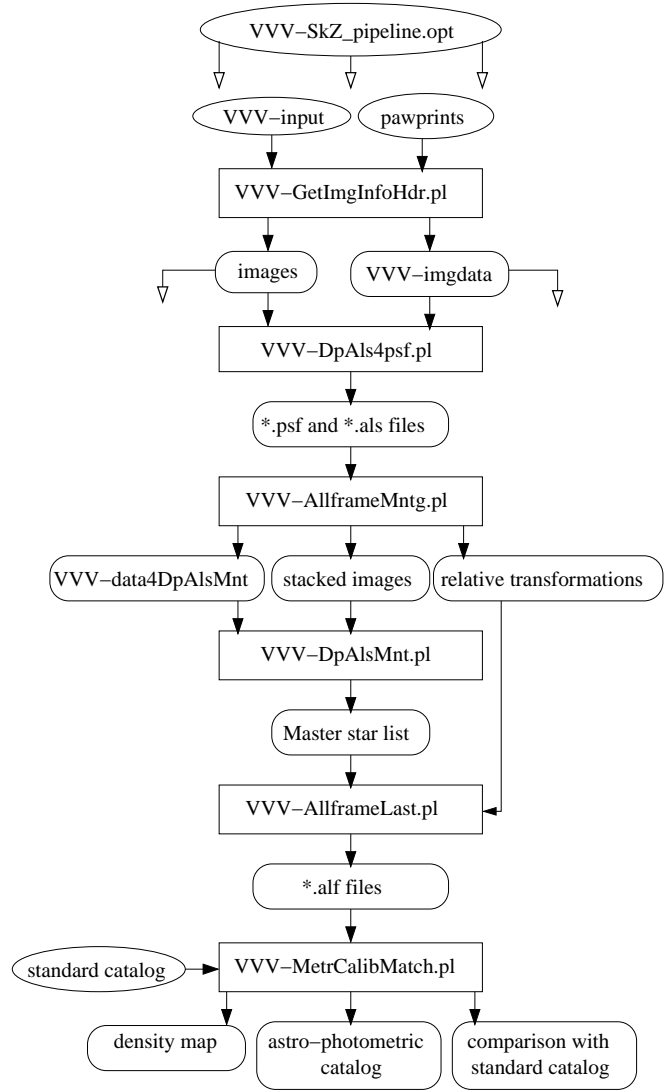}
\caption{Flow chart of the pipeline procedure.
The files the user provides are shown in an elliptical box, the scripts are in  a rectangular box, while the produced files are in a rounded box.
Open arrowheads indicate that the file will be used globally by the following scripts.}\label{fig:flowchart}
\end{figure}

\subsection{Preparing the Input Files}
\label{ss_prep}

The \VSp{} requires three input files in addition to the decompressed pawprints:
\emph{login.cl} (the IRAF configuration file),
\emph{VVV-input} (the input file), and
\emph{\VSp{}.opt} (the configuration file).

The \emph{login.cl} can be simply copied from the IRAF home directory without any change; the pipeline will set by itself the correct type of terminal, restoring the original file at the end.

The input file \emph{VVV-input} must contain three entries per line (see Appendix \ref{ap_input}):
\begin{enumerate}
\item full name of the uncompressed pawprint (e.g., \emph{v20100407\_00619\_st.fits});
\item basename (name without extension) of the output image (e.g., M22-001);
\item space-separated list of chips that have to be extracted (e.g., \textsf{10} or \textsf{10 11}).
\end{enumerate}

The pipeline configuration file \emph{\VSp{}.opt} (see Appendix \ref{ap_opt}) can be used to set a large quantity ($\sim  50$) of options, but all the parameters unspecified in the file will assume their default value.
Hence, its minimum content is related to the catalog of standard stars:
\textsf{stdcat} to specify the path of the catalog;
\textsf{filterstdcat} to indicate the ordered list of passbands;
\textsf{stdposcoomag} to specify the column number of the right ascension (the declination must be in the following column) and of the first magnitude listed for the standard stars (the other magnitudes must be in the following columns, each followed by its error).
Giving a name for the overall photometry with the option \textsf{workname} is preferable, otherwise the default name \textsf{workname=OBJ} will be used.

\subsubsection{Extraction of images and information}

The first step of the procedure is the extraction of the selected frames from the pawprints and of the required information from the headers.
This operation is done by a Perl script that reads its input from the file \emph{VVV-input}.
The script automatically renames the output images.
The file \emph{VVV-imgdata}  stores all the information which will be used later in the processing or can be useful to evaluate the data quality, such as seeing, airmass, sky level, ellipticity, version of the CASU pipeline used and the OB status.

\subsection{PSF calculation}
\label{ss_psfcal}

The PSF is calculated in five steps, increasing in complexity during the first three iterations: in the first step  a purely analytic constant PSF model is used, then a constant look-up table of empirical corrections is added; in the following steps the look-up table is allowed to vary with position in the frame,  quadratically in the last two iterations.
In each step, the pipeline removes each selected PSF-building star that \DAOP{} marks as problematic.
If the PSF calculation should fail, the pipeline changes the analytic function to find the problematic stars, removes them, and then returns to the chosen function.
At the end of the first step, if the number of PSF stars is less than a given value (by default 50), the pipeline re-executes, increasing the initial number.
By default, the VSp uses 400 stars as the initial number of the PSF-building stars (at the end of the first step only the 250 brightest stars are kept), Moffat3.5 as the analytic function, and 8 pixels as the PSF radius \citep[$\sim 2\farcs 7$, where the maximum FWHM seeing allowed for the collection of VVV data is $0\farcs 8$, see][]{Minniti10}).

After each PSF calculation, \ALS{} is run to perform a PSF-fitting photometry of all the stars.
The procedure updates the global source list and the PSF-star list with their improved positions after each run of \ALS{}.
At the beginning of the first two reiterations, the procedure looks for missing stars in the star-subtracted image produced by \ALS{}.
Aperture photometry is subsequently performed on them, after subtracting neighbor stars.
Stars to which this procedure is not applied are processed in the standard way.
\ALS{} is rerun to further improve the coordinates and the number of measurable stars.
This information is used in the following aperture photometry steps, whose results are thus progressively improved.
To reduce the contamination by nearby stars, the PSF is improved using an image where the stars in the proximity of the PSF-building stars are removed using SUBSTAR.
Between the third and fourth step, the user is given the option to visually check the subtraction of the PSF stars and create a list of undetected stars.
The last iteration simply refines the PSF, without adding stars or changing the complexity of the model PSF.
The script gives a well-detailed output including number of stars used, rejected stars, analytic function used, and $\chi^2$ of the PSF provided by \DAOP{}.

\subsection{Creation of the master list of stars and final PSF-fitting photometry}
\label{ss_masterlist}

The master list of stars is generated on a stacked image, created with Montage2.
After a first run of \ALF{} to improve the coordinates of the stars,  the script generates a stacked image from the frames and their star-subtracted versions.
The images in all the available bands are combined, while the combination of different VVV offsets depends on the option activated by the user, following the scheme outlined in Section~\ref{s_general}.
Using the stacked images and the previously generated input file \emph{VVV-data4DpAlsMnt}, the script generates a master list of stars.
This script uses \DAOP{} and \ALS{} in five steps with a procedure similar to the one used for the PSF calculation, where the first step is executed on the stacked star-subtracted images.
The default value of the threshold detection of sources is 4 times the background rms level for the stacked star-subtracted images, and 3 for the normal stacked images.
This master list is then used by \ALF{} to produce the final instrumental PSF-fitting photometry.

\subsection{Final astrometrization, calibration and matching}
\label{ss_metrcalib}

\begin{figure}[ht!]
\centering
\includegraphics[width=\textwidth]{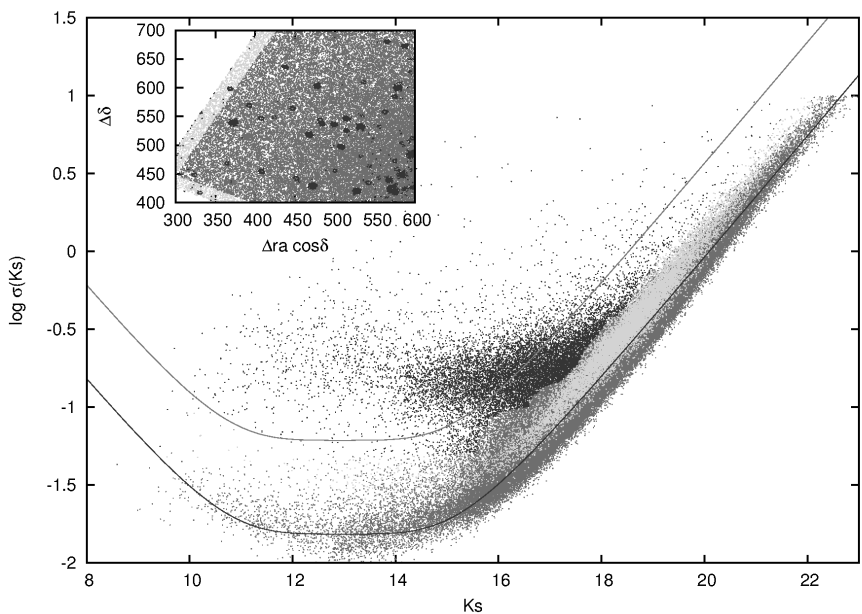}
\caption{Example of the results of the spurious detection cleaning procedure.
We indicate in dark gray the retained sources, in black the rejected sources, in light gray the sources located within 60 pixels from the border.
The main plot shows how the rejected sources are selected in the $(mag;\log{err})$ plane: the black solid line is the fit for the trend of the logarithm of the errors with the magnitude, the gray solid line is the sigma-clipping limit.
Note how the empirical procedure manages to select more sources than sigma-clipping.
The  inserted plot shows an example of the positions of the sources in the relative coordinate system: the rejected ones mainly tend to clump around the position of saturated stars.
The sources along the border present poorer photometry.
} \label{fig:spdetcln}
\end{figure}

The last script initially astrometrizes the \ALF{} output using the WCS of the images, and applies the spurious detection cleaning procedure.
The cleaning process is composed of two independent procedures.
The first procedure is based on the trend of the maxima of the distribution of photometric error for each magnitude  $\sigma_m$.
We empirically found that the distribution of these points can be fitted by a function of the form 
\begin{equation}\label{eq:err_m}
\log{\sigma_m}=a[(.4(m-m_0))^4+b]^{0.25}+c \;.
\end{equation}
The star selection is done rejecting every source with an error greater than \emph{n} times (where \emph{n} is left as an option, with $n=3$ as the default value) the resulting $\sigma_m$ obtained by equation \ref{eq:err_m} (see Figure \ref{fig:spdetcln}, lower plot).
The second procedure is based on both the typical loci of the spurious detections in the magnitude-error diagram, and their characteristic clustering around the saturated stars.
The star selection is performed by an algorithm tuned to minimize the rejection of real stars.
The cleaning process is configurable and reversible, and the positions, magnitude-error and magnitude-$\chi^2$ graphics of both the retained and rejected stars are plotted in an output file ( an example of the results is shown in Figure \ref{fig:spdetcln}).
Analyzing the photometry of several fields, the distribution of the $\Ks{}$ magnitudes of the spurious detections is concentrated mainly around $\Ks{}\approx 15-16$ with a dispersion of 1 magnitude.
A similar peak is sometimes present at $\Ks{}\approx 12-13$.
Even in the other passbands, the distribution is concentrated similarly around magnitude $15-16$.
The color distribution presents a peak, but varies with the field.
In Figure \ref{fig:spdetcln}, the result of the spurious detection selection is shown.
The false sources clearly tend to concentrate in magnitude.
In the inserted plot, we show an example of how the procedure selects the majority of the spurious detections originated by saturated stars.
Around the brightest saturated stars, the spurious detections are also arranged along rays radiating from the center, in a ``cartwheel'' distribution.
As seen in Figure \ref{fig:spdetcln}, the sources lying in the 60 pixel border area have poorer photometry  and this area, covered by only one of the two jitters, is a major source of spurious detections due to its lower signal-to-noise ratio.
For this reason, any source in this 60-pixel frame is rejected, unless this option is disabled.
The pipeline, in addition, rejects all the false detection located in the two empty $60\times 60$ pixel areas in the left-bottom and right-top corners of the images, generated by the jitter pattern.
Similarly it rejects the false detections in the empty areas in chip 1, caused by masking of sensor defects\footnote{\url{http://casu.ast.cam.ac.uk/surveys-projects/vista/technical/known-issues}}.

After this cleaning process, the pipeline calculates the calibration parameters using the standard star catalog, calibrates the photometry, and matches all the data to produce a single output catalog  for all the available bands and offsets.
The calibration procedure considers only standard stars inside a magnitude interval given as input (by default or by the user).
Some standard stars may have visual companions resolved in the VVV, but not in the standard catalog.
If this companion is bright enough, the magnitude in the standard catalog is affected.
The VSp rejects all standard stars with sources in the VVV catalog nearer than a tunable value, whose total light contamination exceeds a  defined threshold.
The default values are a distance of $2\farcs 2$ (tuned assuming 2MASS PSC as the standard star catalog, whose seeing\footnote{\url{http://www.ipac.caltech.edu/2mass/releases/allsky/doc/sec4_4c.html}} varied between $2\farcs 5$ and $3\farcs 4$) and a contamination of 0.03 magnitudes.
The least-squares fitting program assigns a weight-correction factor\footnote{the algorithm is base on a series of five lectures presented at ``V Escola Avancada de Astrofisica'' by Peter B. Stetson \url{http://ned.ipac.caltech.edu/level5/Stetson/Stetson_contents.html} \url{http://www.cadc.hia.nrc.gc.ca/community/STETSON/homogeneous/Techniques/}} to the data to put less weight on the furthest points, instead of doing a sigma-clipping cut.

The calibration is performed twice.
A first calibration is applied to the astrometrized \ALF{} output for the classical correction for zero point and color term \[M_{STD}-m_{VSp}=a_1(J-\Ks{})_{STD}+a_0\;.\]
Since the pipeline uses a local standard system, it does not correct for aperture or atmospheric extinction (both corrections are  constants directly included in the $a_0$).
The second calibration is applied to the output of \DAOM{}  as a zero-point correction, after matching photometry of the same band and the same ``stripe''.
Eventually, the user is allowed to override the determination of $a_0$ and $a_1$ and of the zero-point corrections, imposing her/his preferred values.

%%%%VISTA calibration

The non-equivalence of the two VISTA and 2MASS photometric systems causes non-bijective linear transformation (i.e., stars with the same color could need different magnitude corrections).
This degeneracy is difficult to disentangle because 2MASS PSC and VVV data overlap over a narrow interval of only 2-3 magnitudes.
For this reason, CASU encourages one to calibrate onto the VISTA photometric system, using 2MASS magnitudes to define theoretical VISTA magnitudes, including correction for Galactic extinction.
Of course, the user can directly choose the CASU catalogs of aperture photometry as input reference, thus calibrating directly into the VISTA system while retaining the advantage of the PSF-fitting procedure.
Alternatively, the procedure followed by the VIRCAM pipeline run at CASU to produce the corresponding catalogs can be adopted to calibrate the VSp products onto the VISTA system, feeding the code with a 2MASS input catalog transformed as described in the CASU website.
This flexibility of the VSp is important to guarantee a wide applicability of its products, because the use of the VISTA system, preferable from a theoretical point of view, is not desirable in all astrophysical cases.
For example, many photometric indexes used to study the Galactic globular clusters \citep{Ferraro2006} are not defined in the VISTA system nor, to our knowledge, the transformation of the interstellar absorption among different photometric bands ($A_\lambda/A_V$).

\section{M\,22: a test bed showing the advantages of the pipeline}
\label{s_comp}

\begin{figure}[!ht]
\centering
\includegraphics[width=\textwidth]{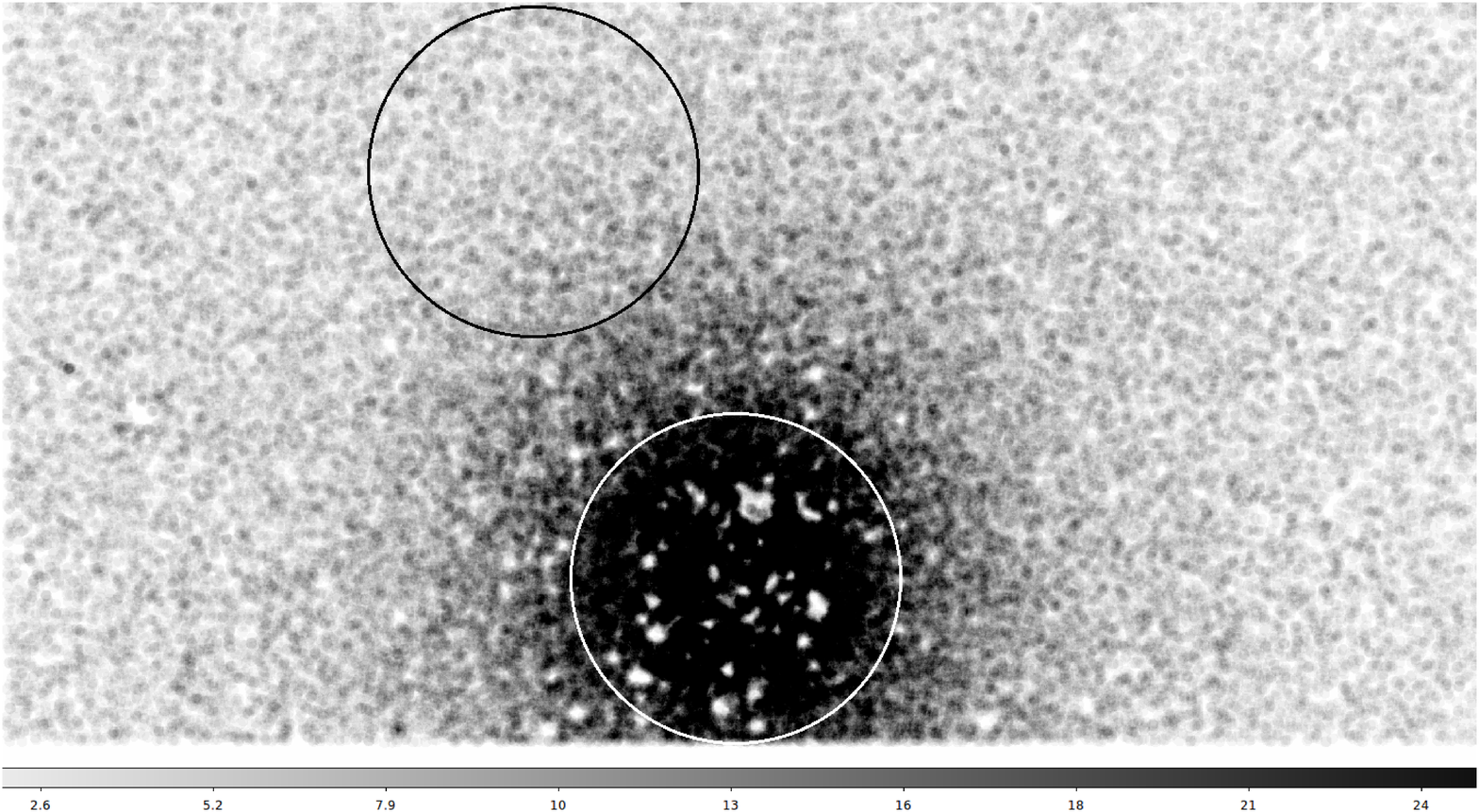}
\includegraphics[width=\textwidth]{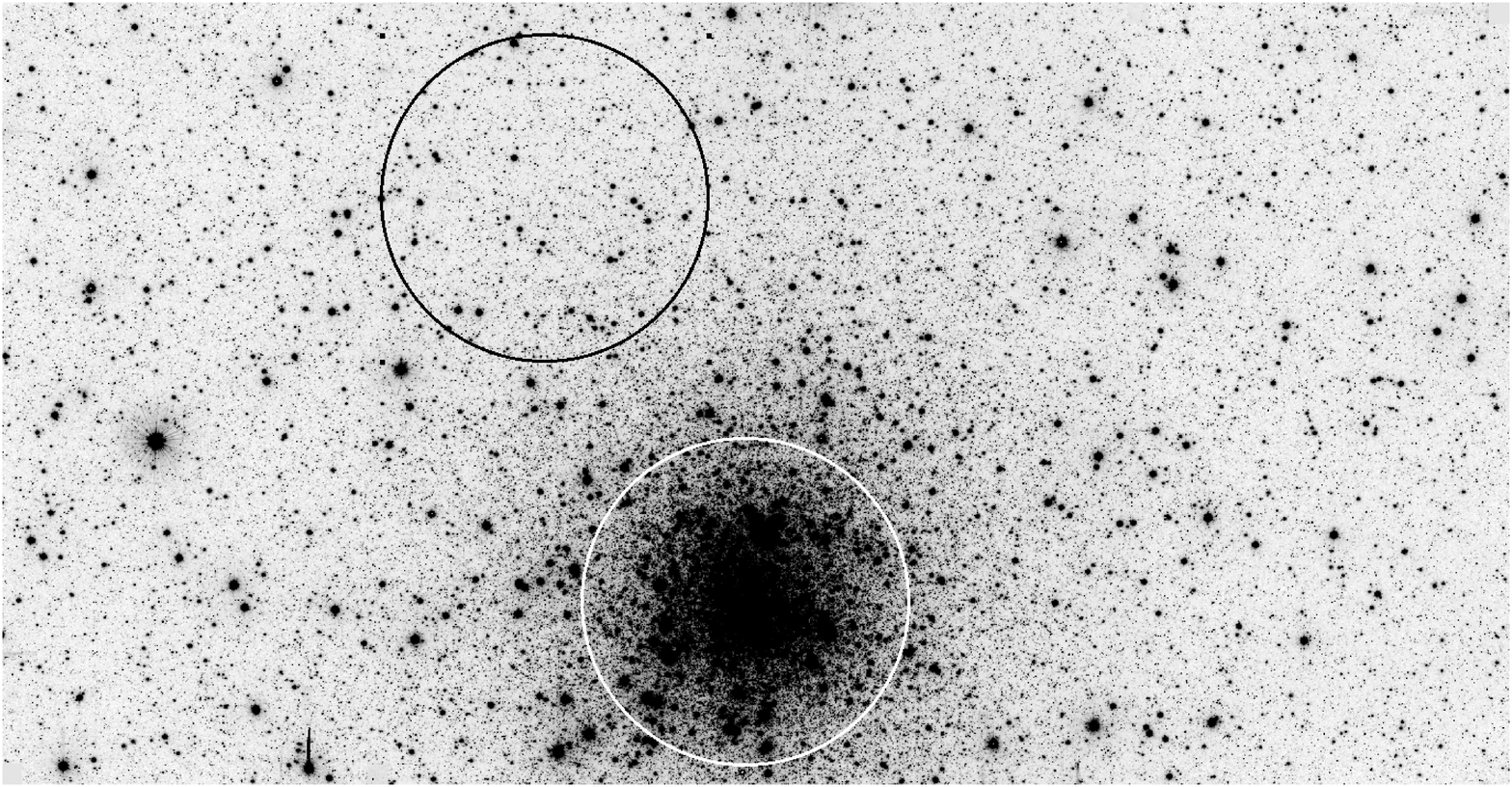}
\caption{Density map produced by the \VSp{} (upper figure) and a whole-field stacked image (lower figure) with  the two subareas used to analyze the luminosity distribution in different condition of crowdedness: the ``central area'' (marked in white) and the ``off-center area'' (marked in black) are shown.}\label{fig:maps}
\end{figure}

In this section, we use the cluster M\,22 (NGC\,6656) as a test bed to compare the VSp products with alternative photometries and codes.
This cluster was chosen  for being a very crowded object in an uncrowded part of the survey.
The test-bed is an area of $\approx 12\arcmin\times24\arcmin$ including the central part of  the cluster M\,22 (see Figure \ref{fig:maps}).
The data comprise, for each of the three offsets covering the area, the three bands  $JHK_\mathrm{s}$ taken in sequence.
We first compare the VSp product with the 2MASS PSC (2M), which is a general reference for near-IR photometry.
The aim is to show the quality of the anchoring of the pipeline photometry to the 2MASS system, and illustrate the brighter limit for good photometry of the VSp products.
In addition, we compare the VSp catalog with two alternative photometric sources: the VVV CASU aperture-photometry catalog (the official catalog of the survey), both the tile and pawprint version (CasuT and CasuP, respectively);  and a PSF-fitting photometry obtained with an updated version of \DoP{} \citep[DoP;][]{DoPhot1989, Alonso2012}.
The CASU catalogs are publicly  available directly at the ESO archive, with no need to download the images.
DoP was written ``to be fast, highly automated, and flexible regarding the choice of PSF'' \citep{DoPhot1993}, therefore it is a valid solution to the problem of extracting  PSF-fitting photometry from a large set of data.

The catalogs were matched with a quite elementary algorithm, which iteratively matches the sources in different catalogs based on both their coordinates and magnitudes simultaneously.
The procedure worked satisfactorily since, while a few percent of unmatched stars is expected, $\approx 90\%$ of the 
sources were matched after the first iteration.
The mean difference of astrometric positions of VSp sources with respect to 2MASS PSC is $-0\farcs 007$ in right ascension and $-0\farcs 001$ in declination, with sample standard deviation (sSD) of $0\farcs 187$ and $0\farcs 180$, respectively.
The differences with the other catalogs are smaller than $0\farcs 05\simeq 0.15$ pixels ($0\farcs 001$ with the DoP catalog) with sSD $\lesssim 0\farcs 11\simeq 0.32$ pixels.
The excellent agreement is not surprising, because all the procedures use the pawprint WCS for astrometrization, which was defined in the CASU pipeline in the 2MASS astrometric system.
Additionally, the images come from the central chips, that observe the part of the field of view less affected by distortion, hence the merging of the different position measurements is less affected by systematic errors.

\begin{figure}[!hp]
\centering
\includegraphics[width=\textwidth]{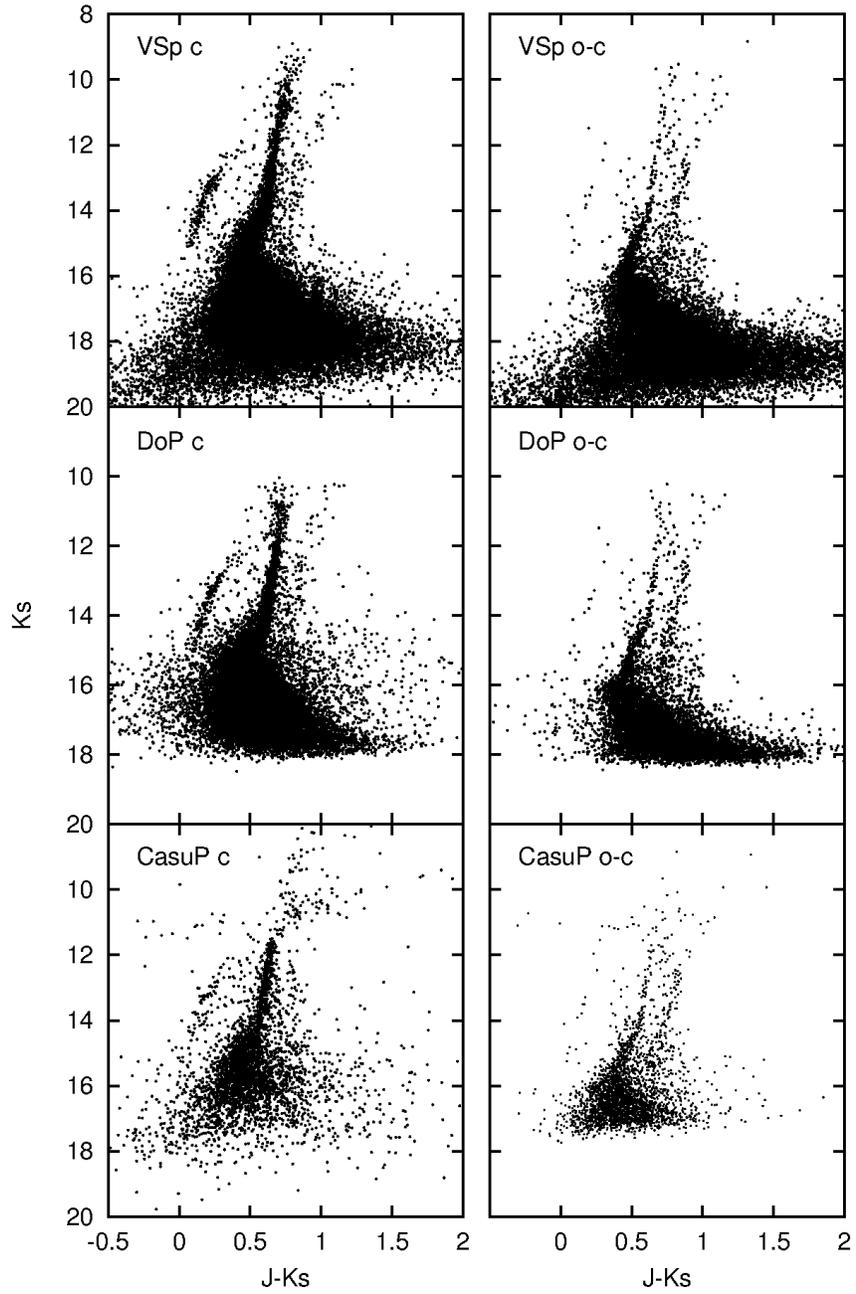}
\caption{Color-magnitude diagrams of the two subareas for the three catalogs based on pawprints: in the left panel the ``central area'' (c), and in the right panel the ``off-center area'' (o-c).}\label{fig:cmdareas}
\end{figure} 

The luminosity distributions of the catalogs is one aspect that must be analyzed with care, because the different crowding in the field influences the source detection.
For this reason, we studied the luminosity distribution in the whole field and in two subareas (see Figure \ref{fig:maps}) of  radius $2\farcm 5$.
The first one was centered on M\,22, and characterized by strong crowding and saturated stars (hereafter the central area), while the second one, located $\sim 7'$ away, was an uncrowded area lacking saturated stars and, consequently, spurious detections (hereafter the off-center area).
In Figure \ref{fig:cmdareas}, we also show the CMDs for the two selected areas for the three pawprint-based catalogs.

\begin{figure}[ht!]
\centering
\includegraphics[scale=.8]{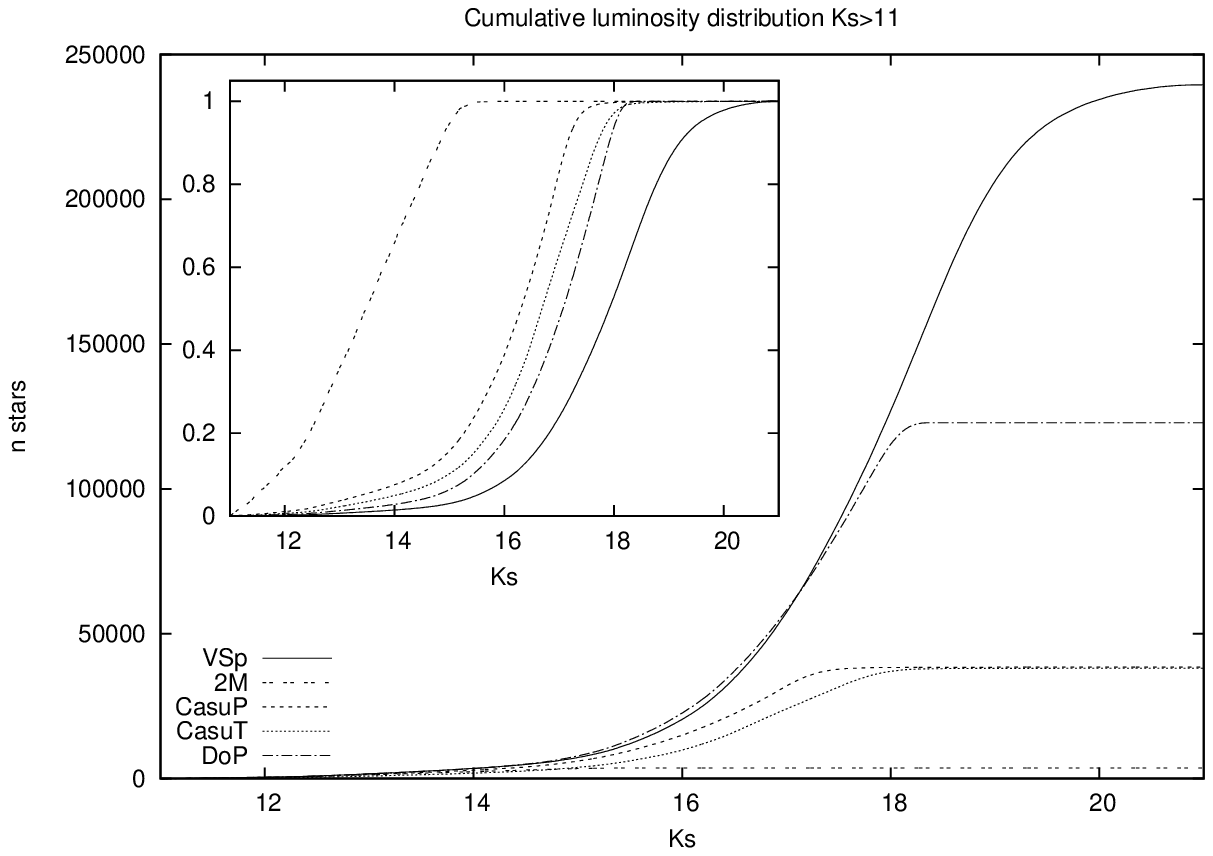}
\includegraphics[scale=.8]{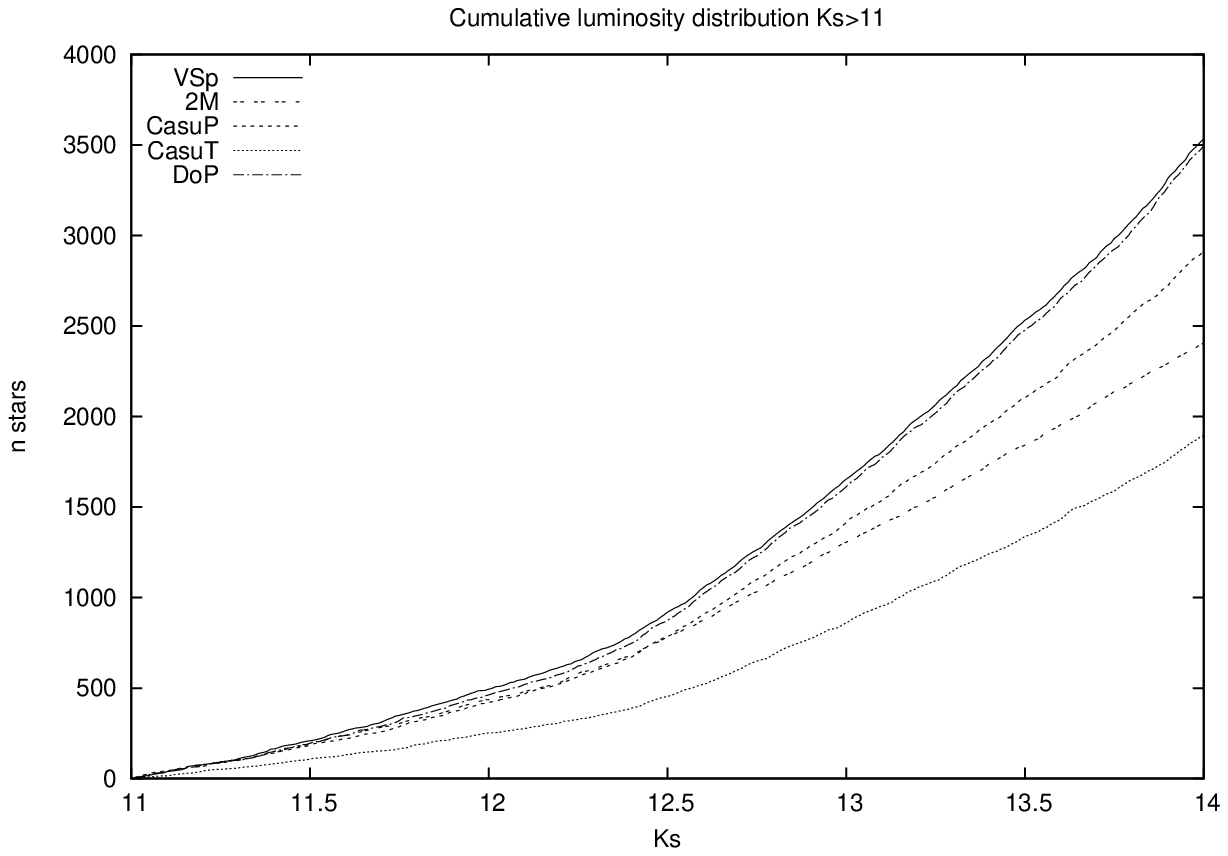}
\caption{Cumulative distribution of the $\Ks{}$ magnitude for $\Ks{}\geq 11$ for the entire field.
The lower panel shows a zoom of the range $\Ks{}=11-14$.
In the in-plot figure the cumulative fraction distribution is showed.} \label{fig:lumdistrC}
\end{figure}

\begin{figure}[ht!]
\centering
\includegraphics[scale=.8]{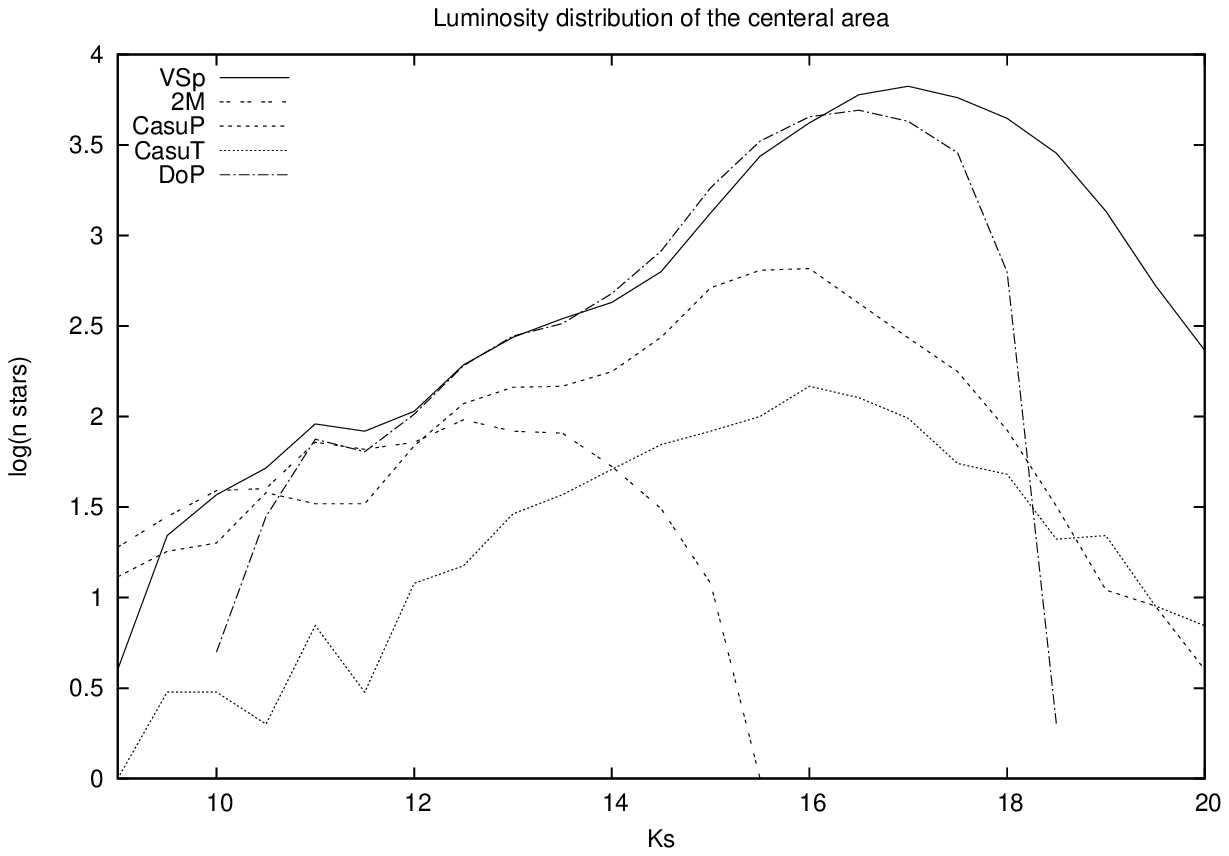}
\includegraphics[scale=.8]{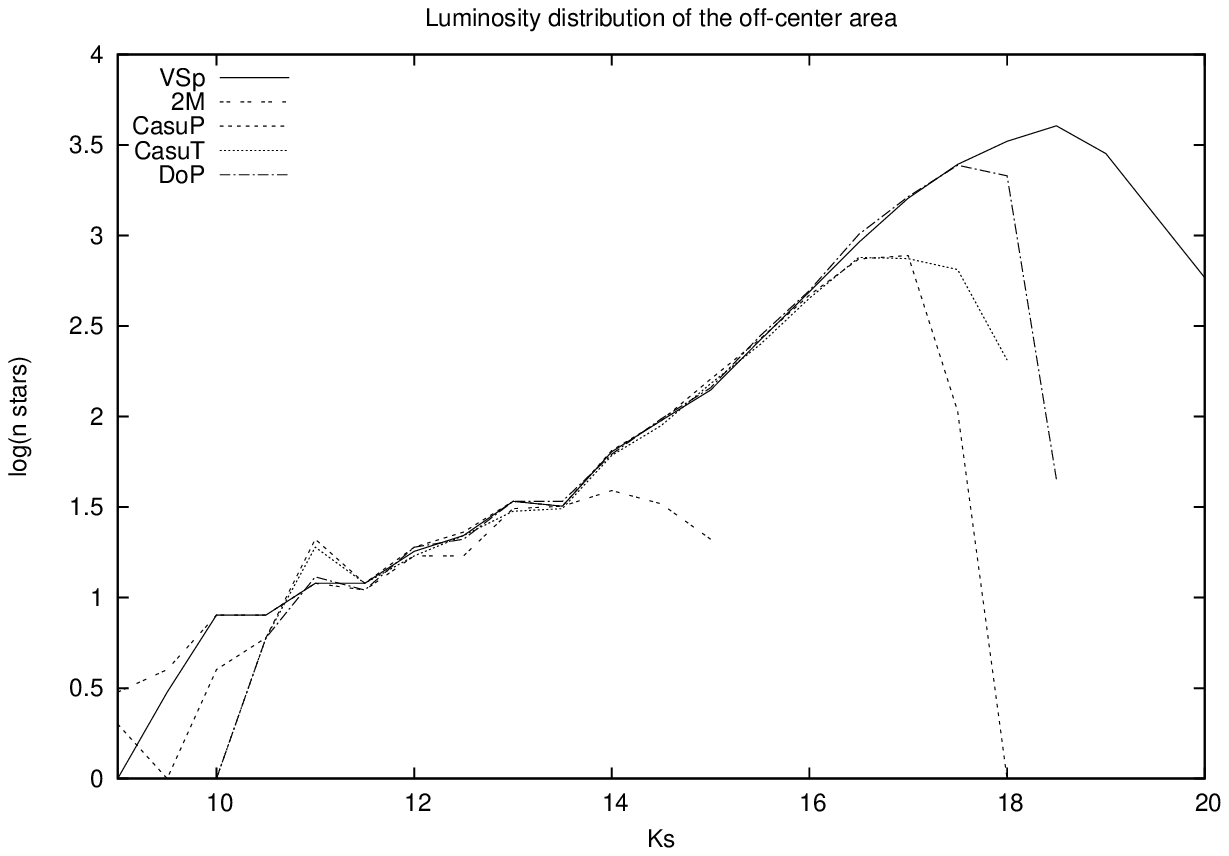}
\caption{Luminosity distribution in the $\Ks{}$ band in the central and off-set areas (upper and lower panel, respectively).} \label{fig:lumdistr25}
\end{figure}

\paragraph{The luminosity distribution.}
The global and fractional cumulative luminosity distributions of sources with magnitude $\Ks{}\geq 11$ for the entire field are shown in Figure \ref{fig:lumdistrC}.
The VSp catalog goes deeper than the others, which are incomplete at $\Ks{}\gtrsim 18$, even if the same threshold for source detection izxs set in the VSp and DoP procedures.
The luminosity distributions of the off-center area (see Figure \ref{fig:lumdistr25}) show how the four procedures detect the same quantity of stars in the range $\Ks{}=11.75-16.25$, matching the luminosity distribution of the 2MASS catalog for $\Ks{}=11.25-13.75$, when the crowding and the incidence of heavily saturated stars is low.
The aperture photometry loses stars for $\Ks{}\gtrsim 16.25$, and only the VSp luminosity distribution  increases down to $\Ks{}=18.75$.
This trend is also clearly visible looking at the CMDs in the right panel of Figure \ref{fig:cmdareas}.

In the central area, instead, the limitations of aperture photometry with respect to  PSF-fitting photometry emerge clearly (CasuP recovers only 40-65\% of the sources with $\Ks{}=12-15$ found by VSp and DoP, and CasuT only 10\%), as seen in the CMDs in the left panel of Figure \ref{fig:cmdareas}.
According to the luminosity distributions, DoP apparently detects more stars with $\Ks{}=14-16$.
As discussed in Section \ref{s_general}, we found that the spurious detections have their main peak in this magnitude range.
The DoP procedure does not clean the spurious detections, and we suspect that this is the source of the increased quantity of apparent source detections.
Further evidence is the ``cartwheel'' distribution of the DoP sources, not detected by VSp, and the presence of more stars with bluer and redder color in the DoP CMD in the left panel of Figure \ref{fig:cmdareas}.

In conclusion, the VSp catalog goes deeper. In addition, it then also returns  more accurate photometric measurements, because the contamination from undetected faint sources is reduced.
In addition, the VSp catalog is less affected by spurious detections, producing a cleaner CMD.
The aperture photometry catalogs are limited with respect to source detection, especially in the most crowded areas.

\begin{figure}[ht!]
\centering
\includegraphics[scale=.8]{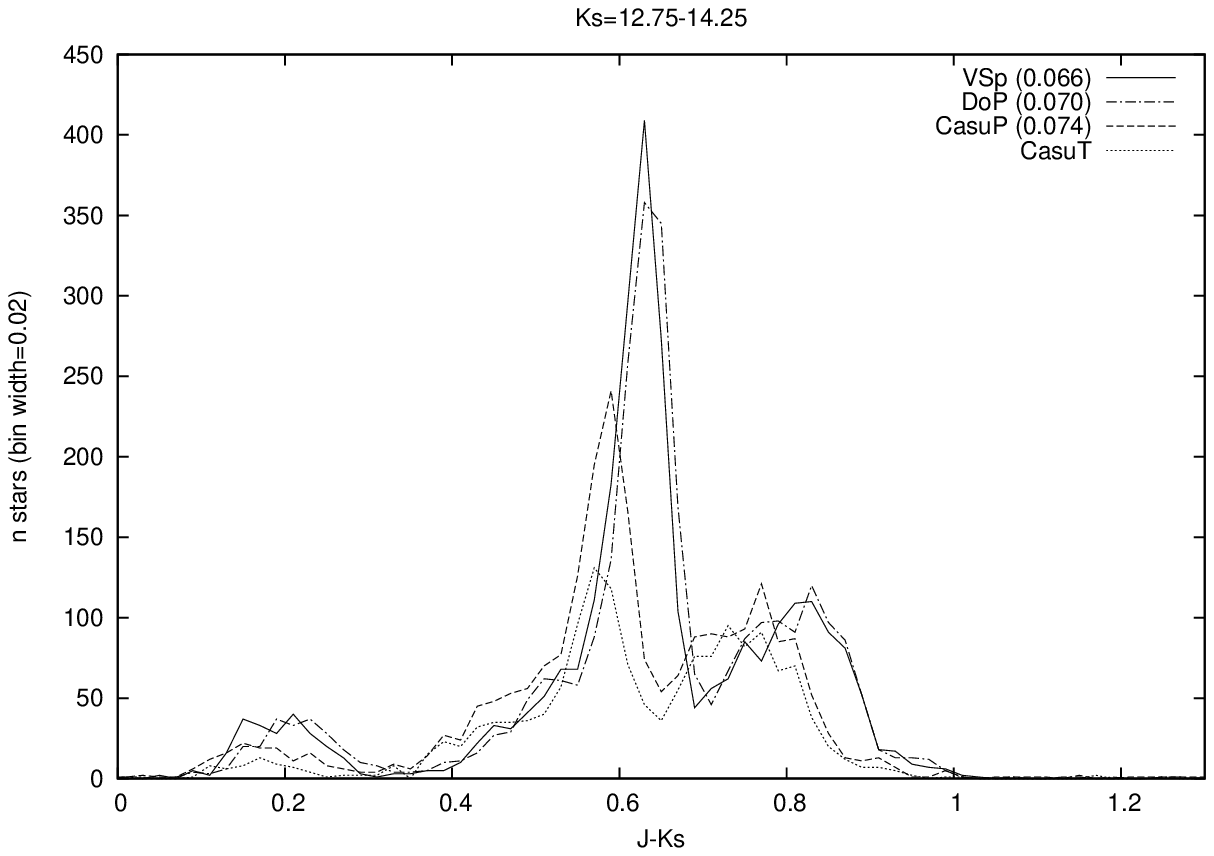}
\includegraphics[scale=.8]{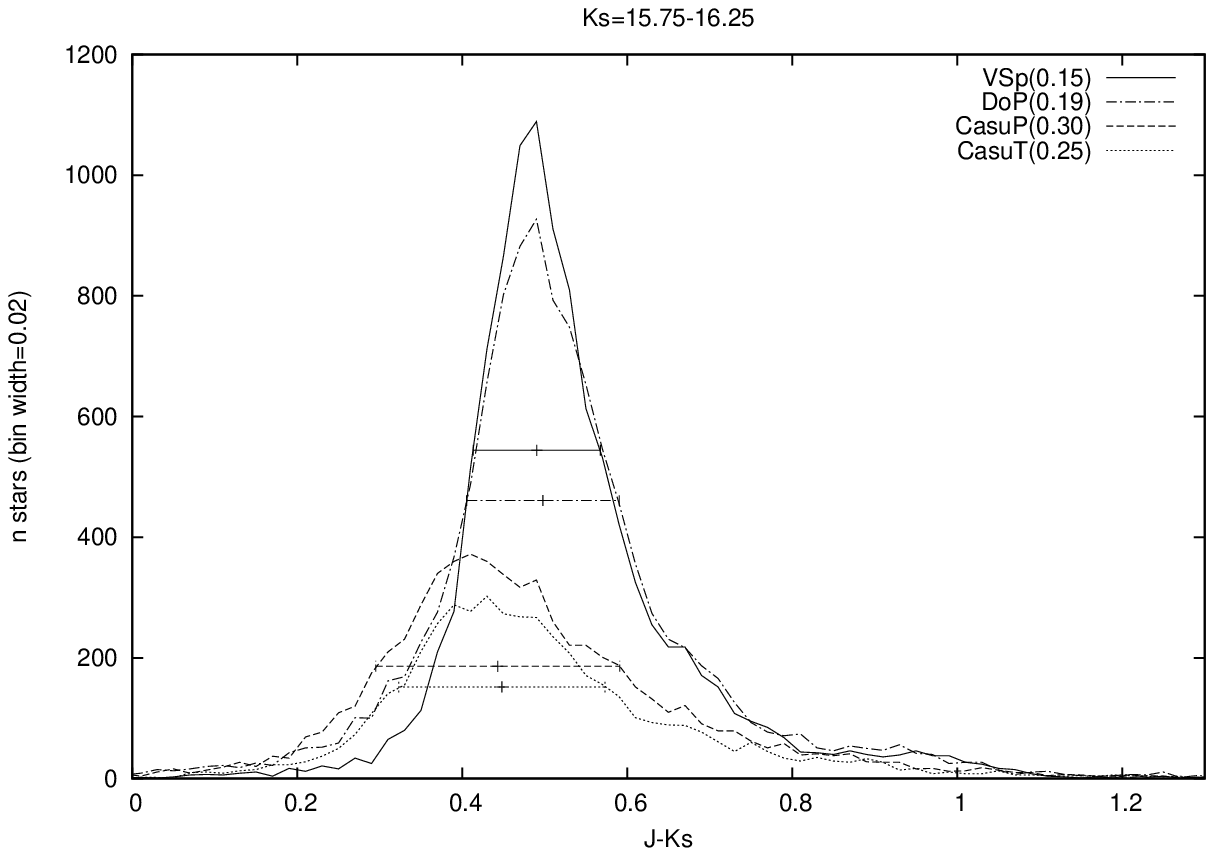}
\caption{Frequency distribution of the color $(J-\Ks{})$ for the four catalogs for stars in the range of $\Ks{}=12.75-14.25$ (upper figure) and  $\Ks{}=15.75-16.75$ (lower figure).
The FWHM of the distributions in color are indicated between parenthesis.}\label{fig:distrJ-Ks}
\end{figure} 

\paragraph{The photometric errors.}
The VSp catalog declares larger errors than the other catalogs for sources with $\Ks{}<16$.
However, this is not a good indication of the precision because some procedures might overestimate or underestimate the true photometric uncertainties.
For this reason, we analyzed the $(J-\Ks{})$ color distribution of stars in three magnitude ranges: $\Ks{}=12.75-14.75$ (lower RGB stars, where the photometry is more accurate; upper panel of Figure \ref{fig:distrJ-Ks}),  $\Ks{}=15.75-16.75$, and $\Ks{}=16.75-17.25$ (MS stars, where the scatter due to photometric errors dominates, but the luminosity distributions are still comparable; lower panel of Figure \ref{fig:distrJ-Ks}).
The observed color distribution is determined by the convolution of the natural spread of the CMD with the photometric errors.
The color spread of the VSp data has the lowest FWHM in all three magnitude ranges, hence they are the least affected by photometric errors.
In the brightest range, the color width of the cluster RGB is $1.06-1.12$ times narrower  in VSp data than in the other two pawprint-based catalogs, while the VSp catalog declares photometric errors $\sim 1.6$ times larger in color.
The disagreement between a narrower dispersion and larger declared photometric errors is not present in the two faintest ranges.

The ratio $\eta=\sigma_{FWHM}/err_{max}$ between the FWHM-derived Gaussian standard deviation (StdDev) and the maximum of the $(J-\Ks{})$ error distribution indicates the goodness of the photometric error estimate.
For each color distribution in the three ranges, the FWHM was calculated and $\sigma_{FWHM}$ was obtained through the inversion of the equation $\mathrm{FWHM} =   2 \sqrt{2 \ln (2) } \; \sigma$.
As stated above, this theoretical StdDev is the convolution of the photometric errors (which depend on the catalog) with the natural spread of the CMD (constant for all catalogs).
Only the pawprint-based catalogs will be analyzed.
In fact, the tile version of the CASU catalog missed most of the stars in the central part of M\,22 and it will be excluded from the following analysis.
VSp data presents a nearly constant $\eta$, slightly increasing at fainter magnitudes (1.63;1.75;1.85 in the three magnitude ranges defined before), while it noticeably decreases for DoP   and for CasuP data (2.71;2.08;1.64 and 2.96;1.63;1.27, respectively).
Thus, $\eta$ is similar in the three catalogs in the faintest two ranges, while in the brighter range it is lower for the VSp data by a factor of 1.66-1.82.
This is caused by the smaller color spread and the higher declared errors for $\Ks{}<16$.

In conclusion, the VSp data present the smallest color spread at any magnitude, therefore the photometric errors are actually smaller.
This improved precision is partially due to the higher detection rate of the faintest sources.
Consequently, its larger declared errors could be an a overestimate, but the constancy with magnitude of the ratio $\eta$, compared to its rapid decrease in the other two catalogs, indicates that the errors are most probably underestimated in the latter.

\paragraph{The photometric differences among catalogs.}

\begin{figure}[ht!]
\centering
\includegraphics[scale=.8]{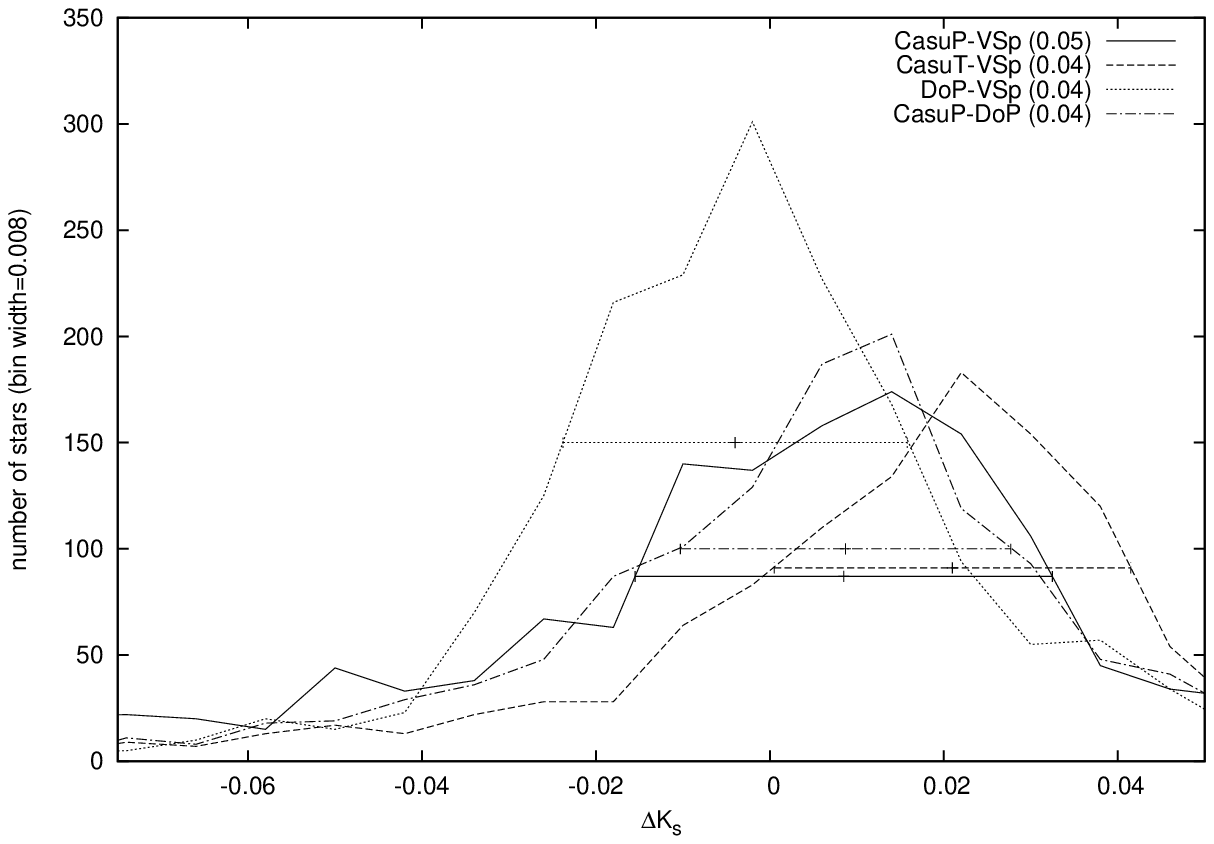}
\includegraphics[scale=.8]{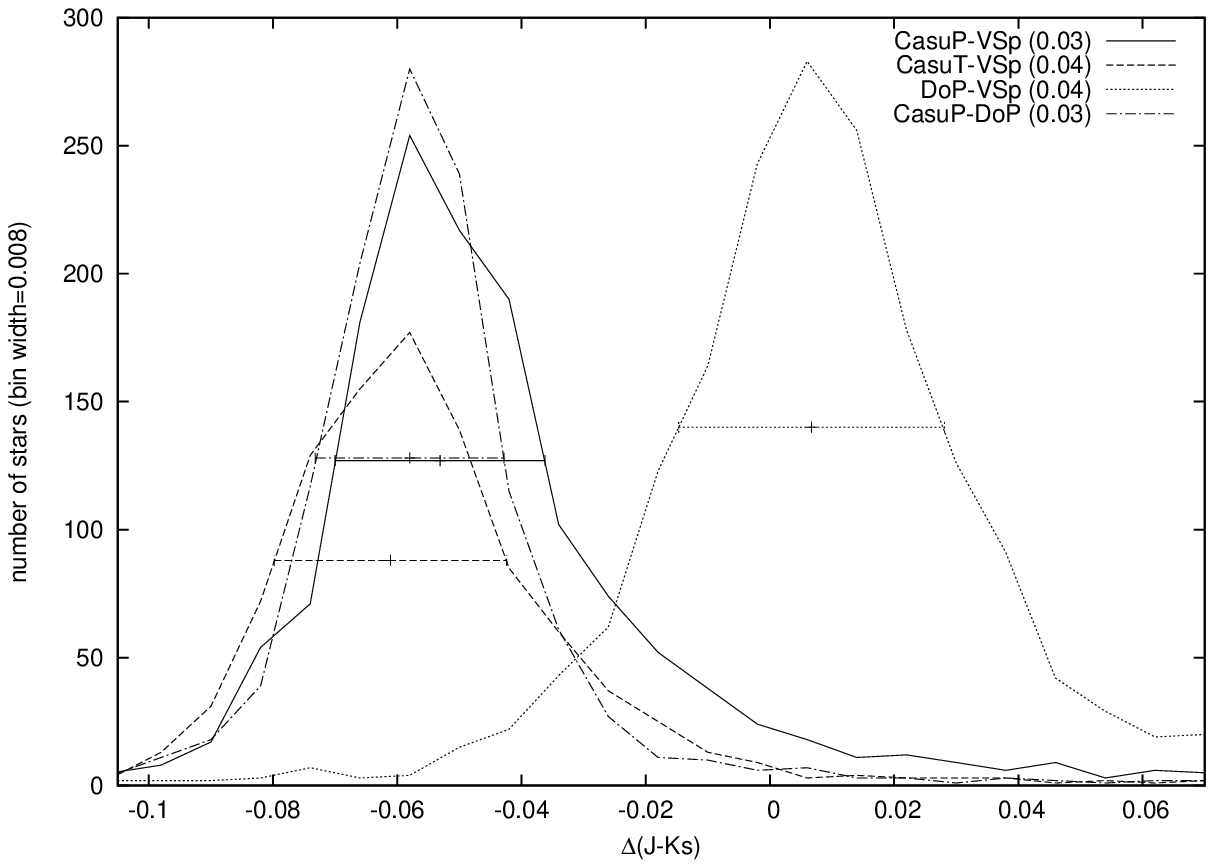}
\caption{Frequency distribution of the differences in $\Ks{}$ (upper figure) and $(J-\Ks{})$ (lower figure) among the catalogs for stars in the range of $\Ks{}=13-14$ (the FWHM of each distribution is indicated in parenthesis).}
\label{fig:distrdiff13-14}
\end{figure}

The frequency distributions of the photometric difference in $\Ks{}$ and in $(J-\Ks{})$  between the catalogs are plotted in Figures \ref{fig:distrdiff13-14} for stars with magnitude $\Ks{}=13-14$.
This range is optimal for this comparison, because it is unaffected by saturation or large photometric errors.
Contrary to the PSF catalogs, the CASU catalogs are calibrated on the VISTA photometric system.
This difference in the photometric system affects mainly only the central position of the distributions of the photometric differences, but not the overall shape, since the considered interval in magnitude is small.

The FWHMs of the distributions are very similar both in magnitude and color.
The $\Ks{}$ differences between the VSp catalog and the other pawprint-based photometries show a plateau in the negative part down to $\sim-0.02\;mag$.
A simple calculation reveals that this is explained by a contamination from sources with $\Delta m\simeq 4.3$~mag, i.e.$\Ks{}\simeq 17.3-18.3$, the magnitude range where the pipeline detects more sources than the other catalogs.
This result underlines the importance of a higher detection rate of faint sources, because the contamination from nearby faint stars introduces errors in the estimated source luminosity.
The distributions of the differences in $(J-\Ks{})$ color are approximately symmetric, but have a narrower dispersion than what is predicted by the photometric errors alone.
It is even narrower than the distribution of the differences in $J$ and $\Ks{}$ magnitude.
This behavior can be explained by a dispersion in $\Ks{}$ due to photometric errors which is stochastic overall, but systematic for a single source, which is thus reduced when the color is calculated.
Such errors can be generated by contamination from  faint sources, variable from source to source, but systematic in all the passbands.
The contamination from nearby undetected sources can worsen the photometric precision by hundredths of a magnitude, even in the brighter part of the photometric range.

\begin{figure}[ht!]
\centering
\includegraphics[]{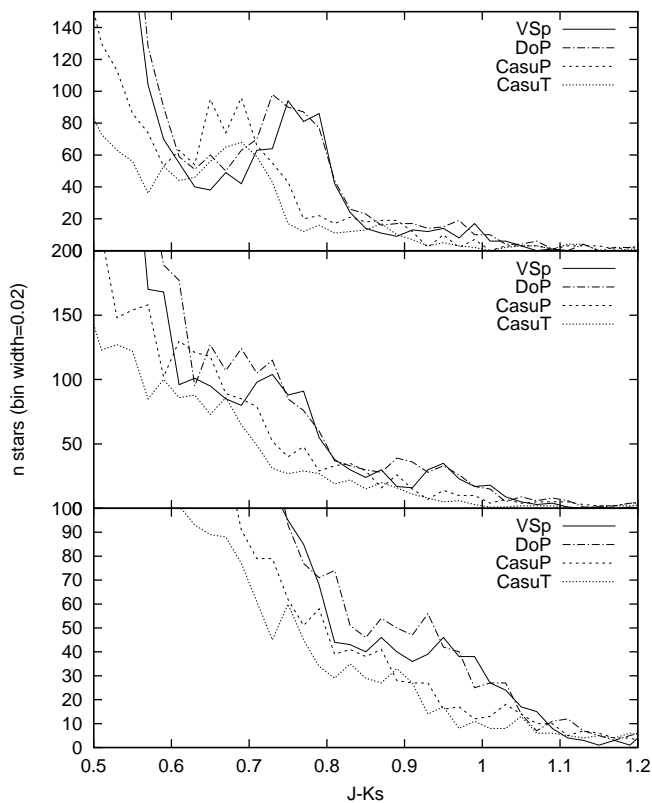}
\caption{$(J-\Ks{})$ color distribution for the four catalogs for stars in the range $\Ks{\,VSp}=14.75-15.25$ (upper figure),  $\Ks{\,VSp}=15.25-15.75$ (central figure) and $\Ks{\,VSp}=15.75-16.25$ (lower figure) }\label{fig:distrJ-Ks_pec}
\end{figure}

\paragraph{The detectability of features in the CMD.}
Another aspect of the photometric accuracy is the feasibility of distinguishing real features of the CMD from the statistical noise.
The $(J-\Ks{})$ color distribution of the field sequences at $(J-\Ks{})=0.75$ and 0.95 (see Figure \ref{fig:cmd}) was analyzed at three different ranges of magnitude,  $\Ks{}=14.75-15.25$, $\Ks{}=15.25-15.75$ and $\Ks{}=15.5-16.25$ (see Figure \ref{fig:distrJ-Ks_pec}), using all four catalogs.
We use the color distribution to evaluate the detectability since its analysis is quantifiable and objective, in contradistinction to an evaluation by eye of the CMD.

At $\Ks{}\approx 15$ the field RGB is easily detectable in all the catalogs, but in the VSp data it stands more prominently above the minimum between it and cluster stars on its bluer side.
At $\Ks{}\approx 15.5$, it is mostly hidden by noise, but the VSp data  present a narrower peak, disentangling it better from the spread of the {M\,22} stars; at $\Ks{}\approx 16$ it is totally blended with the cluster main sequence.
The secondary field sequence is generally undetectable in the CASU catalogs; 
at $\Ks{}\approx 15$ it is sparsely populated (see Figure \ref{fig:cmd}), and barely visible in the color distribution of VSp data.
However, it is detectable at $\Ks{}\approx 15.5$ as a peak  $4.8\sigma$ higher than the background.
The significance of the same peak is lower by about a factor of two ($2.1\sigma$) in DoP data.
At $\Ks{}\approx 16$ the peak is never significant, reaching only $1.7\sigma$ and $1.3\sigma$ above the background for the VSp and DoP catalogs, respectively.

Concurring with the analysis of photometric errors, the narrower spread in color of the VSp data help to spotlight less populated features present in the CMD, improving photometric analysis.

\subsection{Comparison with 2MASS}
\label{ss_comp2MASS}

The 2MASS survey produced an all-sky homogeneous catalog, and it is therefore the reference data-set for the majority of IR photometric studies.
In our case, the comparison with the \VSp{} catalog must take into account that 2MASS data were used as the standard source for the calibration.

\begin{figure}[ht!]
\centering
\includegraphics[scale=.8]{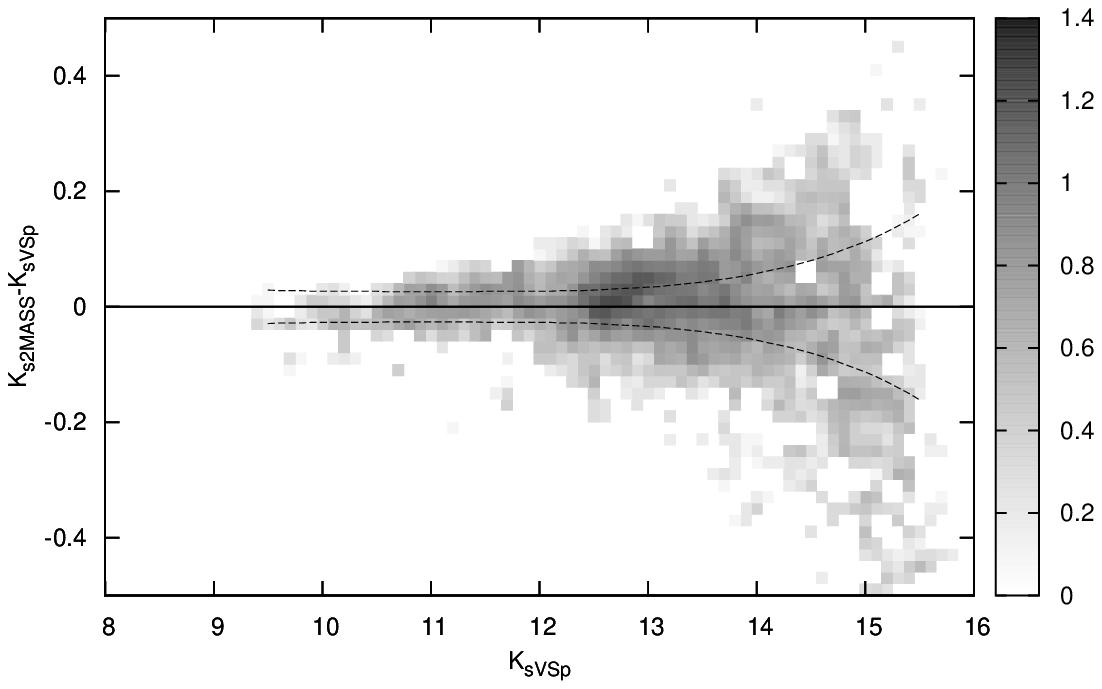}
\includegraphics[scale=.8]{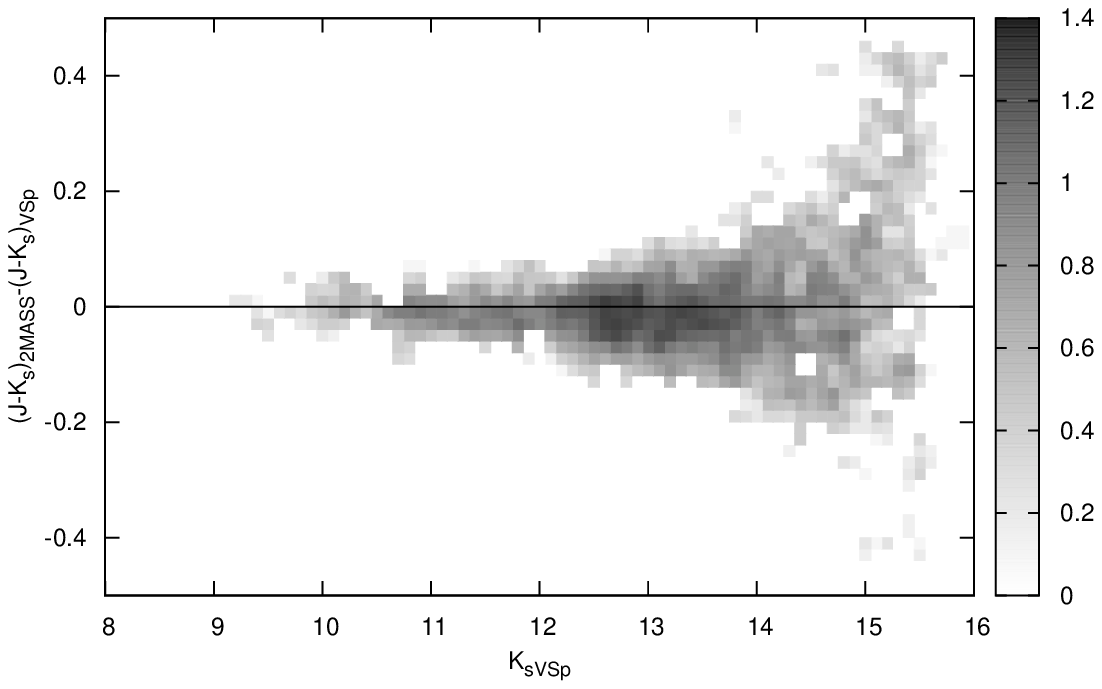}
\caption{Upper figure: density map in logarithmic scale of the photometric differences in $\Ks{}$ between 2MASS and VSp catalogs, as a function of VSp $\Ks{}$ magnitude (the dashed lines indicate the 1-sigma difference).
Lower figure: density map in logarithmic scale of the photometric differences in $(J-\Ks{})$ between 2MASS and VSp catalogs, as a function of VSp $\Ks{}$ magnitude.}\label{fig:2mass}
\end{figure}

As can be seen in Figure \ref{fig:2mass}, VSp catalog shows no significant offset from 2MASS PSC in the range $\Ks{}\approx 9.5-14$.
The partial saturation of stars brighter than $\Ks{}\approx 12.0$ in the VVV survey does not cause any offset in the bright end of the VSp catalog.
For these stars, digital saturation in the A/D converter occurs, but not physical saturation of the potential wells in the detectors, which would cause the  migration of photo-electrons to nearby pixels.
As a consequence, the \DAOP{} suite can fit the non-saturated wings of the PSF and recover the correct stellar magnitude, provided that an optimal \hgd{} value is chosen.
This behavior at the bright end was found in all fields so far tested, and in some cases  it extended  to even brighter magnitudes.
In the case of M\,22, the upper limits of the magnitude range where the VSp photometry is reliable are $J\approx 10.0$, $H\approx 10.0$, and $Ks\approx 9.5$.
In $(J-Ks)$ no offset is observed in the full range $9.5<Ks<15$ (see Figure \ref{fig:2mass}).

In the upper panel of Figure \ref{fig:2mass}, the trend of the error in the magnitude difference is also shown.
This was calculated as the quadratic sum of the photometric errors of the two catalogs.
The observed distribution of ($\Ks{2M}-\Ks{VSp}$) agrees well with the error at all magnitudes.
This was, however, not found in all the fields tested.
Depending on the stellar crowding, the photometry of some 2MASS stars can be contaminated by faint undetected sources, causing  an overestimation in the resulting 2MASS luminosity.
Under these circumstances, the distribution of the magnitude differences is asymmetric, with a predominance of negative values.
Moreover, in crowded fields 2MASS also shows an underestimation of the luminosity of the fainter stars detected, as discussed by \citet{MoniBidin2011}, causing an asymmetry, with a predominance of positive values.
As a consequence, the trend of ($\Ks{2M}-\Ks{VSp}$p) often turns upward in the faintest $1-1.5$ magnitudes of the 2MASS catalog, as also found by \citet{Chene2012}.

\subsection{Comparison with CASU catalog}
\label{ss_compCASU}

The CASU catalog is one of the official public products of the survey, and the astro-photometric catalog of a VVV field that can be obtained with the smallest effort.
The CASU catalogs are offered in two versions: either a single-band catalog for each individual pawprint, or its equivalent for the stacked $1\degr\times1\fdg 5$ tile.
The catalogs are calibrated in the VISTA photometric system, as defined by 2MASS magnitudes and the theoretical transformation equations between the two systems (see Section \ref{ss_metrcalib}).
We have compared our photometry with catalogs obtained from both pawprint (CasuP) and tile (CasuT) versions.
They contain a similar quantity of stars, because the tile version detects more faint sources than the pawprint version, but it loses more stars in the more crowded part of the field (see upper plots in Figure \ref{fig:lumdistr25}).

\begin{figure}[ht!]
\centering
\includegraphics[scale=.8]{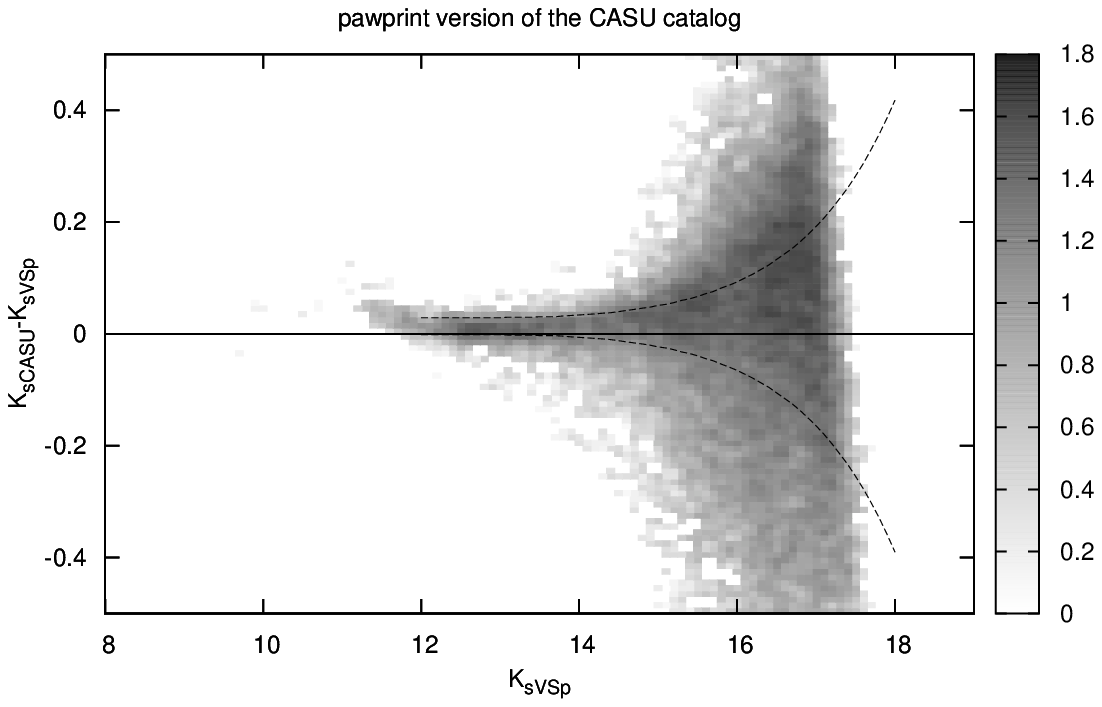}
\includegraphics[scale=.8]{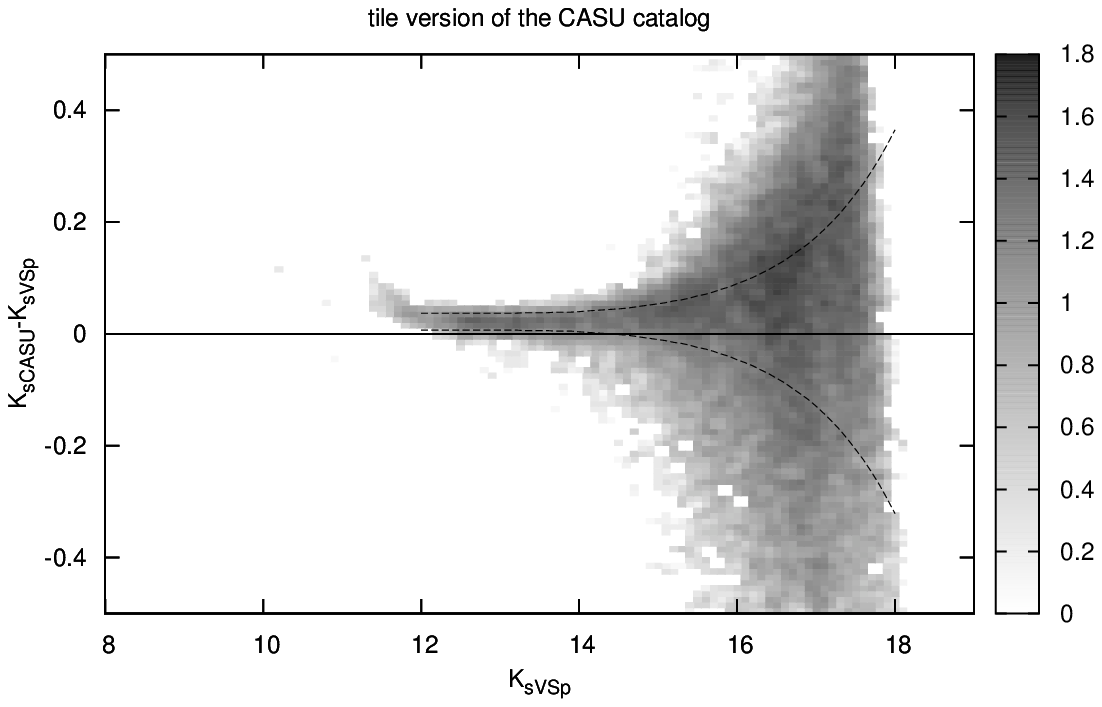}
\caption{Density map in logarithmic scale of the differences in $\Ks{}$ magnitude between CASU (pawprint version in upper figure, tile version in lower figure) and VSp catalogs, as a function of VSp $\Ks{}$ magnitude  (the dashed lines indicate the 1-sigma difference).}\label{fig:casuPT}
\end{figure} 

The photometric comparison is shown in Figure \ref{fig:casuPT}.
The CASU magnitudes clearly deviate in the bright-star regime, as expected for aperture photometry of saturated stars.
This deviation starts at $\Ks{}\approx 12$, $J\approx 12.7$ and $H\approx 12.3$ for both CasuP and CasuT catalogs.
A similar behavior with respect to the 2MASS system for stars close to the saturation limit is also shown in Fig. 2 of \citet{Gonzalez2011} and in the VMC survey paper by \citet{Cioni2011}.
As expected, the catalogs have a small zero-point offset with respect to the VSp catalog, not being in the same photometric system.
However, and quite unexpectedly, the two catalogs do not present the same exact offset, as can be seen in Figure \ref{fig:distrdiff13-14}.
The mean magnitude difference is $\Delta\Ks{}=0.014mag$ for CasuP and $\Delta\Ks{}=0.022mag$ for CasuT, similar to the value $\Ks{VVV}-\Ks{2M}=0.028$ found by \citet{Gonzalez2011} for tile b282.
This means that the two CASU catalogs have a magnitude difference of $\Delta\Ks{}=0.008$, explicable as a systematic error in their zero-point determination.
The average difference in color is $\Delta(J-\Ks{})=-0.058$\,mag for both catalogs, very  similar to the offset found by \citet{Gonzalez2011} between the 2MASS PSC and tile catalog for tile b282.
Such magnitude differences are greater than what is predicted by the CASU transformation  between 2MASS and VISTA systems.

\subsection{Comparison with \DoP{}}
\label{ss_compDoP}

As a last evaluation of the photometric quality of the pipeline products, we compare the VSp catalog with an alternative PSF-fitting photometry code.
Since the pipeline primarily aims to reduce the idle time and user interaction, the comparison catalog was based on an updated version of \DoP{} \citep{DoPhot1989,Alonso2012}, a widely-used, fast, user-friendly and mostly automatic code.
\DoP{} needed 11 minutes of preparation, while the photometry was obtained in 24 minutes.
The final step needed an additional 12 minutes.
The whole photometric procedure took $\sim 47$ min on a MacBookPro with CPU 2.53GHz Intel Core2Duo.
\DoP{} was used with the threshold for source detection equal to the mean sky level plus three times the background signal-to-noise ratio.

\begin{figure}[ht!]
\centering
\includegraphics[scale=.8]{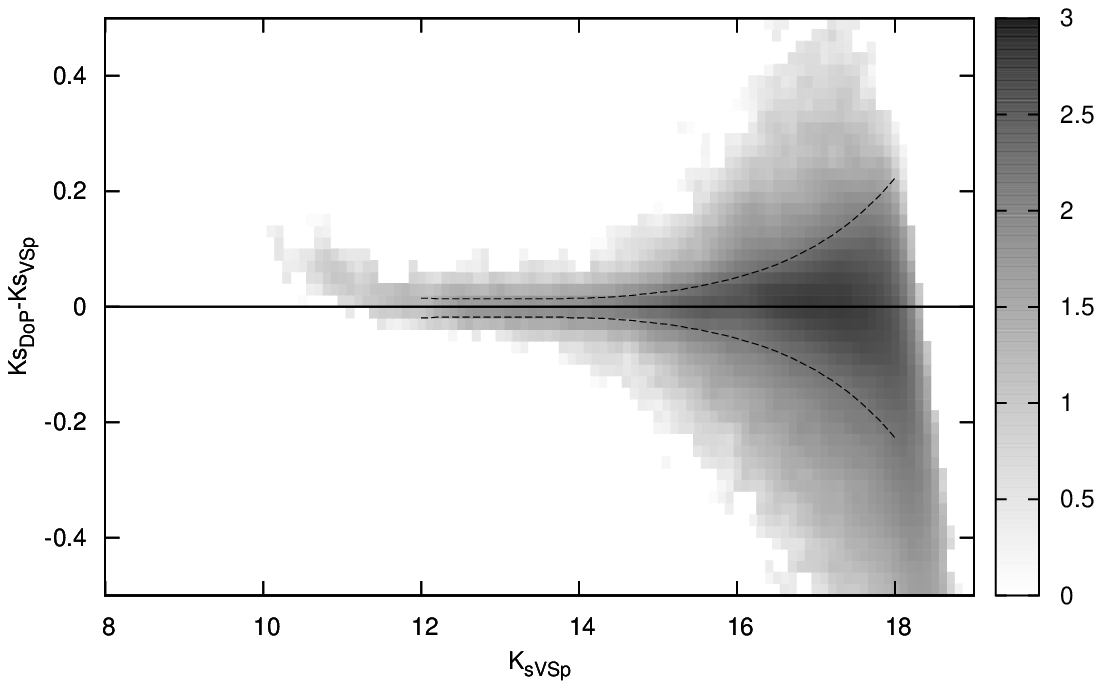}
\includegraphics[scale=.8]{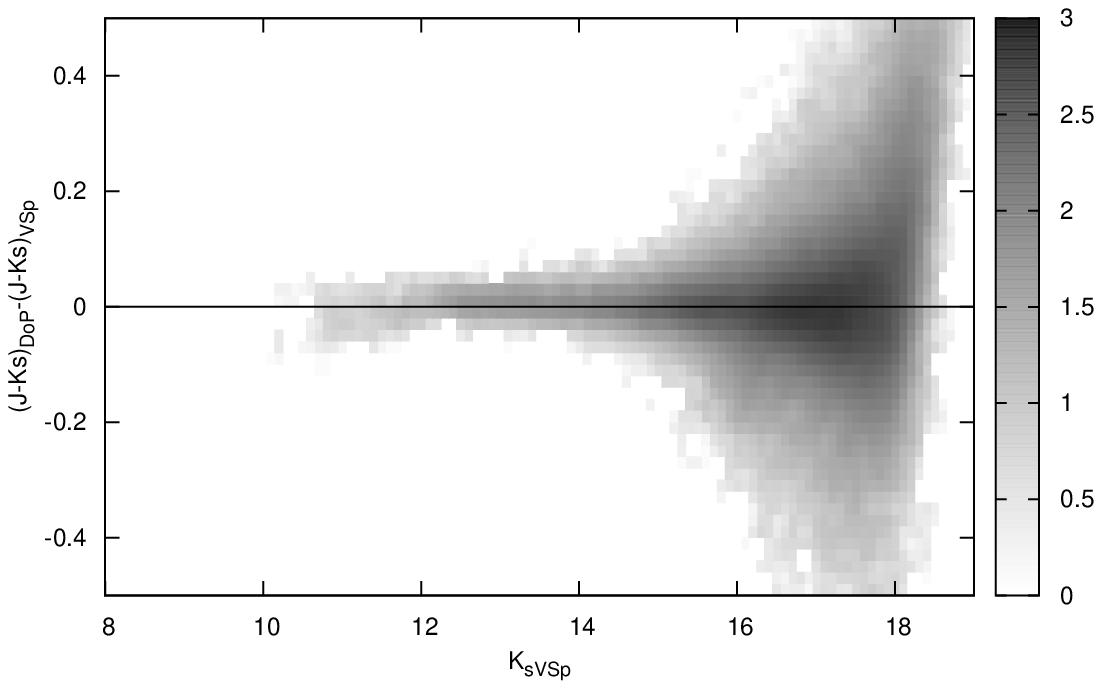}
\caption{Upper figure: density map in logarithmic scale of the photometric differences in $\Ks{}$ between DoP and VSp catalogs, as a function of VSp $\Ks{}$ magnitude (the dashed line indicates the 1-sigma difference).
Lower figure: density map in logarithmic scale of the photometric differences in $(J-\Ks{})$ between DoP and VSp catalogs, as a function of VSp $\Ks{}$ magnitude.}\label{fig:DoP}
\end{figure} 

The match with the DoP catalog presented some problems, since sources with $\Ks{\,DoP}=14.5-17$ were matched with sources brighter and fainter by $1-2$mag. A similar mismatch was present between DoP and CasuP catalogs.
These sources are probably spurious detections, given the magnitude range (see former discussion) and the peculiarity in their position distribution, tending to cluster around the position of saturated stars, usually assuming a ``cartwheel'' distribution.
We found that a $5\sigma$-clipping rejection in the matching procedure resolved the problem.

DoP was originally programmed to mask the saturated stars and their surroundings \citep{DoPhot1993}.
This is a problem in many applications, both because many faint stars are lost in crowded regions, and because many important objects are found in the weakly saturated but still well photometered regime of VVV data, as discussed in Section \ref{s_intro}.
As observed in Section~\ref{ss_comp2MASS}, the \VSp{} can correctly work up to about two-three magnitudes brighter than the saturation limit.
We therefore removed the masking of bright stars in the DoP procedure \citep{Alonso2012}.
This change, unfortunately, also causes the introduction of several spurious detections around saturated stars which are not removed in the procedure.

The photometric comparison (shown in Figure \ref{fig:DoP}) shows that the DoP magnitudes deviate in the bright-star regime.
Hence, even removing the masks, the magnitudes of the saturated stars are not recovered by DoP.
The photometric offsets for fainter stars between the two PSF-fitting photometries are minimal: $\Ks{\,DoP}-\Ks{\,VSp}\approx -0.002$ and $(J-\Ks{})_{DoP}-(J-\Ks{})_{VSp}\approx 0.006$.

\begin{figure}[ht!]
\centering
\includegraphics[width=\textwidth]{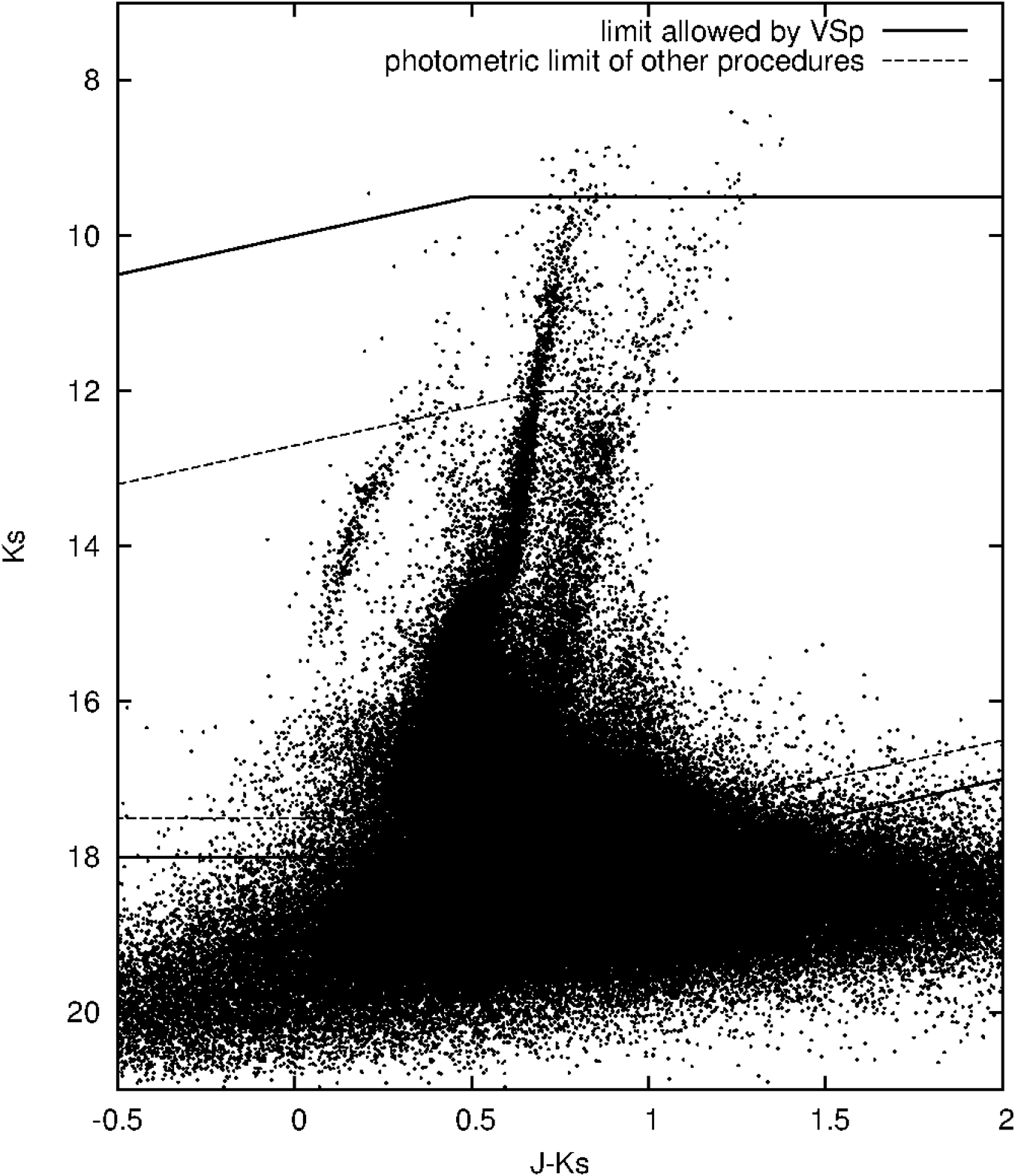}
\caption{Color-magnitude diagram of M\,22 overplotted with the photometrically reliable area allowed by the VSp catalog (within solid lines), and the photometric limits of other catalogs (within dotted line).
VSp permits to use a magnitude range 50\% larger.}\label{fig:cmd}
\end{figure} 

\subsection{\VSp{} performance}
\label{ss_perf}

Here, we discuss the typical processing time for an object.
The code was run on an Intel(R) QuadCore(TM) i7 CPU 3.33GHz with normal (not high-performance) hard disks.
The \DAOP{} suite was compiled in 32bit.

The PSFs of the nine images were calculated in $\sim 2.5$ hours, with a $\chi^2$ that varies from 0.013 to 0.024 (average value of 0.017), and using from 75 to 163 PSF-building stars for each image.
The PSF-building stars have magnitudes $J\approx 12.8-13.4$, $H\approx 12.4-13.0$ and $\Ks{}\approx 12.2-12.7$.
The final cleaned VSp catalog of $\sim 242,000$ stars was obtained in about 10 hours.
We reran the final script in order  to select the optimal ranges of magnitude for the calibration.
The final number of stars reduced per hour is $\approx 24,000$.
Using two additional epochs in  $K_\mathrm{s}$, the PSFs were calculated in $\sim 4$ hours, while the final cleaned catalog of $\sim 245,000$ stars was obtained in about 13 hours, with the number of stars reduced per hour of $\approx 19,000$.
The average error of the color term of the calibration is $\approx 0.011$ (with a scatter of $0.005$).

\section{Conclusions}

In this paper, we have presented the astro-photometric  \VSp{} (VSp), a \DAOP{}+\ALF{} procedure optimized to work on ``VISTA Variables in the V\'ia L\'actea'' (VVV) ESO Public Survey data, a large-area survey covering the Galactic Bulge and southern disk.
We demonstrated that, with very little effort, the user can  obtain accurate results over a photometric range larger than provided by using other solutions such as the CASU catalogs or other automatic PSF-fitting photometry programs like \DoP{} (DoP).

VSp is the only one of the four solutions analyzed in this paper that minimizes the effects of saturation; it provides a deep astro-photometric catalog, reliable more than 2 magnitudes brighter than the saturation limit (see Figures  \ref{fig:2mass} and \ref{fig:cmd}).
At variance with other solutions, VSp permits the study of interesting bright objects, and use techniques that rely on the brightest part of the CMD (like the determination of the metallicity with Calcium II Triplet equivalent width or the slope of the upper RGB) using VVV data.
We showed that, when  2MASS PSC is used as a reference for calibration, VSp produces data well anchored to its photometric system, thanks to its larger photometric range.
On the contrary, other solutions have to rely on stars with $J>12.7$ and $\Ks{}>12$ (for bulge fields, fainter in the disk), where the 2MASS photometry presents deviations that can introduce non-negligible shifts.
Systematic offsets are present also in the CASU catalogs, even if they are anchored to the VISTA system.
In fact, we evidenced shifts in $\Ks{}$ larger than those predicted by CASU transformations between 2MASS and VISTA system.

We showed that VSp yields more detections in crowded fields.
The presence of spurious detections, which adds noise to the CMD and ruins the match with other photometries, is efficaciously handled, removing them.
The VSp detects more faint sources, thus reducing their contamination in the photometry of the brighter sources.
This contamination can worsen the photometric precision by some hundredths of a magnitude even in the brighter end of the VVV photometric range.
These results allow one to produce a cleaner CMD, permitting a better analysis of the chosen target.
Consequently, VSp data produce CMDs with narrower color dispersion, optimal to isolate and spot faint structures.

This pipeline permits the user to obtain a high-level PSF-fitting photometry with all the accuracy of the \DAOP{} suite, with the advantage of easy use and minimum idle time.
In fact, the \VSp{} requires only the preparation of the data and possibly some additional iterations of the calibration procedure.
In addition, VSp provides an adequate number of options to tune all the procedures according to the needs, and provides a set of output files to check in detail the result.
The VSp pipeline will be distributed upon request, contacting the first author at \email{fmauro@astroudec.cl}

\section*{Acknowledgements}

We gratefully thank Peter B. Stetson for the \DAOP{} suite and his explanations.
We thank Karen Kinemuchi for her help in the testing and checking of the reliability of the original pipeline and her suggestions for the paper.
Dante Minniti, the PI of the VVV project, is  warmly thanked for instigating this very important survey.
We gratefully acknowledge support from the Chilean {\sl Centro de Astrof\'\i sica} FONDAP No.15010003 and the Chilean Centro de Excelencia en Astrof\'\i sica y Tecnolog\'\i as Afines (CATA) BASAL PFB-06/2007.
CMB received support from The Milky Way Millennium Nucleus.
ANC received support from Comite Mixto ESO-Gobierno de Chile and GEMINI-CONICYT No. 32110005.
JAG was also supported by the Chilean Ministry for the Economy, Development, and Tourism's Programa Iniciativa Cient\'ifica Milenio through grant P07-021-F, awarded to The Milky Way Millennium Nucleus, by Proyecto Fondecyt Regular 1110326, by Proyecto Fondecyt Postdoctoral 3130552, and by Anillos ACT-86.
JB is supported by FONDECYT  No. 1120601 and by the Ministry for the Economy, Development, and Tourism's Programa Inicativa Cient\'{i}fica Milenio through grant P07-021-F, awarded to The Milky Way Millennium Nucleus.
This publication makes use of data products from the Two Micron All Sky Survey, which is a joint project of the University of Massachusetts and the Infrared Processing and Analysis Center/California Institute of Technology, funded by the National Aeronautics and Space Administration and the National Science Foundation.
IRAF is distributed by the National Optical Astronomy Observatories, which are operated by the Association of Universities for Research in Astronomy, Inc., under cooperative agreement with the National Science Foundation.

\begin{appendices}

\section{Mandatory External Programs}
\label{ap_progr}

The pipeline calls some external programs, all freely distributed with no charge,  that must be previously installed on the work machine:\\
\texttt{\DAOP{} suite}.
The pipeline uses the stand-alone version to obtain the photometry; the user has to ask it directly to P. B. Stetson and compile it.\\
\texttt{Perl}.
The pipeline is composed mainly by scripts written in Perl that operate as a front-end for the \DAOP{} suite.
The interpreter is usually already installed in every Unix/Linux distribution (type \textsl{perl -v} on a command line to find out which version)\footnote{\url{http://www.perl.org/get.html}}.\\
\texttt{WCStools}.
The pipeline uses part of the programs in this package (like sky2xy, xy2sky, cphdr) to manage the WCS transformations.
It is downloadable from the Smithsonian Astrophysical Observatory site\footnote{\url{http://tdc-www.cfa.harvard.edu/wcstools/}}.\\
\texttt{NOAO IRAF}.
The pipeline needs \textsc{iraf} \citep{iraf}, v2.13 or following, mainly to extract the images from the pawprints and the header informations, and to create the density map in fits format; moreover the check of the PSF is operated with an interactive iraf task.
It is downloadable from NOAO IRAF site\footnote{\url{http://www.iraf.net}}.\\
\texttt{gnuplot}.
The pipeline uses this program to generate the plots needed to check the outputs.
It is usually included in software available for every Linux distribution, but also downloadable from its site\footnote{\url{http://www.gnuplot.info/}}.\\
\texttt{ds9}.
This program is only needed if the user want to use the interactive iraf task to check the PSF calculation.
It is downloadable from the Smithsonian Astrophysical Observatory site\footnote{\url{http://hea-www.harvard.edu/RD/ds9/}}.\\
\texttt{system programs}.
The pipeline uses several system programs (like date, pwd, tar, bzip2, wc, tail, cut, sort,...).
They are usually already installed in every Unix/Linux distribution.
\\~\\
\texttt{NASA HEASARC imcopy} is not used by the pipeline, but the user needs it to decompress the VVV data available at the Vista Science Archive\footnote{\url{http://horus.roe.ac.uk/vsa/}} (VSA).
It is downloadable from the NASA\footnote{ \url{http://heasarc.gsfc.nasa.gov/docs/software/fitsio/cexamples.html#imcopy}} or WFCAM Science Archive\footnote{ \url{http://surveys.roe.ac.uk/wsa/qa.html#compress}} site.

\section{\emph{VVV-input}}
\label{ap_input}

Example of \emph{VVV-input} input file.
In this case the extraction of only one chip per pawprint was necessary.
\begin{verbatim}
v20100407_00619_st.fits  M22-01 10
v20100407_00621_st.fits  M22-02 11
v20100407_00623_st.fits  M22-03 11
v20100407_00631_st.fits  M22-04 10
v20100407_00633_st.fits  M22-05 11
v20100407_00635_st.fits  M22-06 11
v20100407_00643_st.fits  M22-07 10
v20100407_00645_st.fits  M22-08 11
v20100407_00647_st.fits  M22-09 11
v20100825_00508_st.fits  M22-10 10
v20100825_00510_st.fits  M22-11 11
v20100825_00512_st.fits  M22-12 11
v20100826_00420_st.fits  M22-13 10
v20100826_00422_st.fits  M22-14 11
v20100826_00424_st.fits  M22-15 11
\end{verbatim}

\section{\emph{\VSp{}.opt}}
\label{ap_opt}

Example of \emph{\VSp{}.opt} option file.
Only the mandatory options are listed.
\begin{verbatim}
workname=M22
stdcat=../2MASSforVVV.dat
stdcat=../2MASSforb242.dat
filterstdcat=J,H,Ks
stdposcoomag=1,3
magcut=11.5:13.5,11.0:13,10.5:12.5
\end{verbatim}

\section{Format of the final catalog}
\label{ap_Ctlg}

The catalogs have a three-line header.
The first line gives the coordinates of the four corner of the field in the internal coordinate system.
The number of the stars and of the passbands in the catalog are indicated in the second line.
In the third line, the equatorial coordinates of the zero point of the internal coordinate system and the mean angle between the x axis of the images and the right ascension axis, both in radiant and degree units, are given.

The data for each star follow the header.
They consist of:
\begin{description}
\item[Star ID] in the form: number of tile (1-396) - number of stripe in the tile (1-8) - unique source ID in the stripe of 7 ciphers.
\item[Equatorial coordinates] of the star.
\item[Internal coordinates] in arcseconds from the zero point given in the header.
\item[Magnitudes and associated errors] for each passband in the order $JH\Ks{}YZ$
\end{description}

Example of a final catalog.
{\footnotesize
\begin{verbatim}
#    1.449 598.415    1200.803 1.551    1498.941 603.678    306.326 1201.217
#nstars= 178092  nmag= 3
#ra0(J2000)= 278.8988800     dec0(J2000)= -24.0254100    angle  -0.4601670 -26.3656280
24250000001  278.8993207  -23.8591836     1.449  598.415  18.6468   0.0758  17.8891   0.1218  ...
24250000002  278.9003358  -23.8595092     4.787  597.243  17.4458   0.0321  16.9051   0.0488  ...
24250000003  278.9005007  -23.8583633     5.329  601.368  17.3438   0.0219  17.0691   0.0561  ...
24250000004  278.9014003  -23.8555253     8.287  611.585  17.5538   0.0368  16.8541   0.0405  ...
24250000005  278.9015098  -23.8558219     8.647  610.517  17.4228   0.0372  16.6371   0.0394  ...
24250000006  278.9016773  -23.8583997     9.198  601.237  18.8868   0.1269  18.0951   0.1460  ...
24250000007  278.9019067  -23.8558033     9.952  610.584  19.0598   0.1164  18.6451   0.2118  ...
[...]

\end{verbatim}
 }
 
\end{appendices}

\end{document}